\documentclass[aps,twocolumn]{revtex4-2}

\usepackage{myPackage}
\usepackage{myxrhyper}
\usepackage{dcolumn}%

\myexternaldocument[sm-]{build}{sm}

\begin{document}

\title{Beyond Group Delay: Fundamental Limits of Achromatic Metasurfaces}

\author{Roman Buisine}
\thanks{Equal contribution}
\affiliation{Université Côte d'Azur, CNRS-CRHEA, Rue Bernard Gregory, Valbonne, 06650, France.}%

\author{Nicolas Kossowski}
\thanks{Equal contribution}
\email{nicolas.kossowski@crhea.cnrs.fr}
\affiliation{Université Côte d'Azur, CNRS-CRHEA, Rue Bernard Gregory, Valbonne, 06650, France.}%

\author{Adelin Patoux}
\affiliation{Université Côte d'Azur, CNRS-CRHEA, Rue Bernard Gregory, Valbonne, 06650, France.}%

\author{Patrice Genevet}
\affiliation{Université Côte d'Azur, CNRS-CRHEA, Rue Bernard Gregory, Valbonne, 06650, France.}%
\affiliation{Department of Physics, Colorado School of Mines, 1523 Illinois St., Golden, CO, 80401, USA. }%

\author{Rémi Colom}
\affiliation{Université Côte d'Azur, CNRS-CRHEA, Rue Bernard Gregory, Valbonne, 06650, France.}%

\author{Samira Khadir}
\email{samira.khadir@crhea.cnrs.fr}
\affiliation{Université Côte d'Azur, CNRS-CRHEA, Rue Bernard Gregory, Valbonne, 06650, France.}%

\begin{abstract}
       Achromaticity is one of the most demanding functionalities achievable with a metasurface because it requires reproducing the same optical response over a finite spectral bandwidth rather than at a single wavelength. Although previous studies established aperture–bandwidth limits based on temporal-delay considerations, the role of optical performance on these limits has remained unresolved. Here, we develop a rigorous theoretical framework that explicitly highlights how the optical performance sets the true aperture-bandwidth limit of single interface achromatic metasurfaces, quantifying  the fundamental limits through the Strehl ratio. We identified two distinct operating regimes. In the first, the achromaticity conditions are exactly satisfied, guaranteeing diffraction-limited performance across the entire bandwidth. Surprisingly, the maximum achievable aperture in this regime is essentially independent of bandwidth and is instead governed by the ability of practical meta-atoms to simultaneously realize the required phase and dispersion. In the second regime, relaxing the required optical performance allows larger apertures at the cost of a fundamental trade-off between aperture, bandwidth, and optical performance. Numerical validation using a large library of high-transmission \ce{GaN} meta-atoms confirms the theory and reveals the transition between these regimes. Our framework establishes the fundamental limits of achromatic metasurfaces, providing practical design rules for balancing aperture, bandwidth, and optical performance of achromatic metasurfaces.
\end{abstract}

\maketitle


\section{Introduction}
Optical metasurfaces have emerged over the last decade as a powerful platform for manipulating electromagnetic wavefronts using arrays of subwavelength scatterers (meta-atoms) \cite{yu_light_2011,Genevet_17}. By tailoring the geometry of individual meta-atoms, these ultra-thin optical components can locally engineer the phase, amplitude and polarization of light, enabling a broad range of functionalities, including beam steering, holography, structured-light generation, and flat lenses \cite{chen_review_2016}. Due to their compactness, low weight, and compatibility with semiconductor fabrication technologies, metasurfaces are widely regarded as promising candidates to replace conventional bulk optics in future integrated optical systems\cite{zhao_review_2022, kuznetsov_roadmap_2024, brongersma_second_2025}.

For many practical applications, however, monochromatic operation is insufficient. Imaging systems, augmented and virtual reality displays, microscopy, spectroscopy, among other applications, require optical components capable of maintaining the same functionality over a broad spectral bandwidth \cite{chen_flat_2020}. This has stimulated considerable efforts toward the development of achromatic metasurfaces such as metalenses \cite{aieta_multiwavelength_2015, shrestha_broadband_2018, chang_achromatic_2024, engelberg_nature_2025, fan_broadband_2019, lin_achromatic_2019, wang_broadband_2018}, beam deflectors \cite{jia_achromatic_2021} and other flat optical components based on dispersion-engineered dielectric meta-atoms \cite{wang_visible-frequency_2016}. Despite these remarkable advances, the dimensions of experimentally demonstrated single-layer achromatic metasurfaces remain surprisingly small, typically limited to diameters ranging from a few tens to a few hundreds of micrometers in the visible or infrared ranges, suggesting that their scalability is fundamentally more challenging than that of conventional monochromatic metasurfaces \cite{chen_broadband_2018, shrestha_broadband_2018, wang_broadband_2018, chen_broadband_2019, hu_asymptotic_2023, chang_achromatic_2024}.

The origin of this difficulty is deeply rooted in the physics of wavefront control. In a monochromatic metasurface, only the optical phase at a single wavelength must be prescribed. Since the phase is periodic modulo $2\pi$, the required phase delay can always be wrapped into a single optical cycle. Consequently, every meta-atom only needs to provide an optical path difference smaller than one wavelength. Achromatic metasurfaces pose a fundamentally different challenge. Preserving the same optical functionality over a finite spectral bandwidth no longer requires controlling only the optical phase at a single wavelength, but also its spectral evolution \cite{chen_flat_2020}. Equivalently, the meta-atoms must simultaneously provide the prescribed phase delay, group delay, group-delay dispersion, and, for sufficiently broad bandwidths, higher-order dispersion terms \cite{chen_broadband_2018}. Unlike monochromatic phase control, these temporal quantities cannot be reduced by phase wrapping. Instead, the required group delay is directly determined by the optical path difference to be compensated across the device and therefore increases with the metasurface aperture. As a result, while a monochromatic meta-atom typically needs to delay light by no more than one optical period, an achromatic metalens of practical dimensions may require differential delays corresponding to hundreds or even thousands of optical cycles between its center and edge. This enormous difference in the required delay raises the question: can a optical element, with a thickness of the order of the wavelength, accumulate such large temporal delays while simultaneously preserving the desired wavefront with high optical fidelity?

This fundamental limitation has motivated several theoretical efforts aimed at establishing the maximum achievable size of achromatic metasurfaces \cite{shrestha_broadband_2018, presutti_focusing_2020, engelberg_achromatic_2021, liang_high_2019, arbabi_controlling_2017}. Among the first analyses, \cite{shrestha_broadband_2018} derived analytical scaling laws relating the aperture of achromatic metalenses to their operational bandwidth by considering the group delay required to compensate for chromatic dispersion across the device. Their work established the now widely accepted aperture--bandwidth trade-off and provided valuable design guidelines by showing that broadband achromatic metasurfaces cannot be scaled arbitrarily while satisfying the required temporal-delay profile. Identifying that the generation of large group delay is the bottleneck to realize large achromatic metasurfaces, \cite{presutti_focusing_2020} revisited this problem from a more fundamental perspective studying the group delay bounds using concepts previously employed to realize slow-light buffers \cite{baba_slow_2008,tucker_slow-light_2005,miller_fundamental_2007}. While generating large group delay is necessary to realize large achromatic metasurfaces, it is not a sufficient condition as extra constraints exist regarding the phase shift and group delay dispersion. This raises a subsequent question of whether or not metasurfaces realized using these concepts can genuinely eliminate all forms of chromatic aberrations. 

In this work, we develop a rigorous theoretical framework to quantify the fundamental limits of single-interface effective-index achromatic metasurfaces while explicitly incorporating optical performance into the analysis. 
To assess the practical performance of achromatic devices, we evaluate the generated wavefront using the Strehl ratio instead of relying solely on phase or group-delay matching. This metric directly quantifies the fidelity of the optical function, making it particularly relevant for imaging and wavefront-shaping applications. This approach naturally extends previous delay-based analyses by introducing optical performance as an additional fundamental dimension of the problem.
Our analysis reveals that the fundamental limitations of achromatic metasurfaces are governed by two distinct operational regimes. In the first regime, corresponding to perfect achromaticity, all achromaticity conditions are exactly satisfied, \textit{i.e.} the generated phase profile matches perfectly the target phase profile. Surprisingly, we demonstrate that the maximum achievable size in this regime is essentially independent of the operating bandwidth. However, the corresponding constraints are so stringent that they can only be satisfied by extremely small single-interface metasurfaces. In the second regime, which we refer to as quasi-achromaticity, chromatic aberrations are partially tolerated, and the Strehl ratio becomes smaller than unity. In this regime, a fundamental trade-off emerges between aperture, bandwidth, and optical performance. Relaxing the Strehl ratio allows moderately larger devices to be realized, but our analysis shows that the achievable dimensions remain severely constrained for broadband operation. Beyond providing a more rigorous interpretation of previous size-bandwidth scaling laws, our framework offers a practical design tool that enables metasurface designers to qualitatively estimate the range of device dimensions that can be achieved for a given bandwidth and required optical quality.

The paper is organized as follows. Section~\ref{sec:perfect_achromaticity} establishes the achromaticity conditions and derives rigorous bounds for perfectly achromatic metasurfaces. Section~\ref{sec:quasi_achromaticity} extends the analysis to the quasi-achromatic regime by introducing a generalized framework that relates chromatic aberrations to the average Strehl ratio through a Taylor expansion of the spectral phase. Section~\ref{sec:dataset} presents the numerical dataset used to validate the theoretical predictions. Section~\ref{sec:validation} applies the proposed framework to practical achromatic metalenses and discusses the resulting aperture--bandwidth--performance trade-offs. Finally, section~\ref{sec:discussion} discusses the general implications of our results.

\section{Theory}
\subsection{Perfectly Achromatic Metasurfaces}
\label{sec:perfect_achromaticity}

\begin{figure*}
    \centering
    \includegraphics[width=\textwidth]{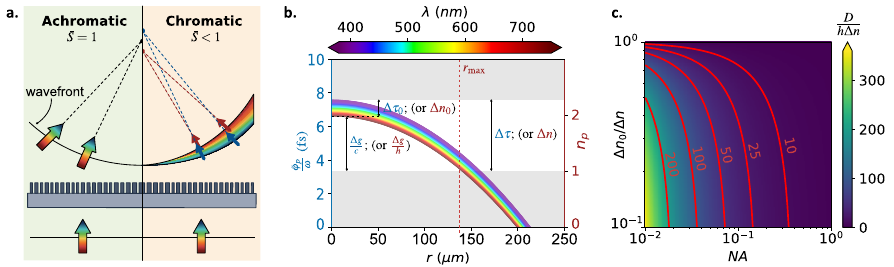}
    \caption[Bounds of Perfectly Achromaticity Metasurfaces]{\textbf{Bounds of perfectly achromaticity metasurfaces.} \textbf{a.} Schematic representation of a chromatic and achromatic wavefront shaping by a metalens with dispersive meta-atoms. Plane waves with different frequencies are normally incident on the metalens. In the chromatic case, the wavefronts at each frequency are different, leading to different focal points, while in the achromatic case, all wavefronts are identical. \textbf{b.} Evolution of the phase delay and effective refractive index $n_p$ of a perfectly achromatic metalens in function on the position $r$ and wavelength $\lambda$. The white areas delimit the total dispersion $\Delta \tau$ of the achromatic subset. \textbf{c.} Maximum achievable diameter of an achromatic metalens $D$, made of highly transmissive truncated pillars, as function of the numerical aperture NA and variation of refractive index of the reference meta-atoms $\Delta n_0$ with regards to the available index variation $\Delta n$.}
	\label{fig:figure_0}
\end{figure*}

\subsubsection{Achromaticity Condition}
In the most general case, the optical function of the wavefront shaping metasurface is defined by the \textit{target phase profile},
 \begin{equation}
     \varphi(\bm{r}, \omega) = \frac{\omega}{c} g(\bm{r}),
     \label{eq:target_phase_profile}
 \end{equation}
where $\omega$ is the operating angular frequency, $c$ the speed of light in free space, and $g(\bm{r})$ the optical path difference to compensate at the position $\bm{r}$ in order to realize the desired phase profile. Typically, the function $g(\bm{r})$ is often determined via ray optics, leading to the convention $g(0) = 0$. For example, for a metalens, with focal length $f$, $g(\bm{r}) = f - \sqrt{f^2 + \abs{\bm{r}}^2}$ and for a 1D deflector along the x-dimension only at angle $\alpha$, $g(x) = x \sin\alpha$.

To realize an achromatic metasurface over a bandwidth $\Delta\omega$, each meta-atom $p$, positioned at $\bm{r}_p$ along the metasurface, should compensate for the phase $\varphi(\bm{r}_p, \omega)$ at each frequency $\omega$ of the bandwidth. However, according to the generalized law of refraction \cite{yu_light_2011}, the wavefront of the wave exiting the metasurface is a function of the gradient of the phase profile $\nabla\varphi$, meaning only the relative phase between the elements of the metasurface is important. We choose a common reference to be the meta-atom placed at the center of the metasurface, \textit{i.e.} at $\bm{r}_p = 0$, and we index it with the subscript $p=0$. Noting $\tilde{\varphi}_p$ the \textit{phase dispersion} of the $p$-th meta-atom of the metasurface, the generalized Snell law translates to,
\begin{equation}
        \tilde{\varphi}_p(\omega) - \tilde{\varphi}_0(\omega) = \frac{\omega}{c} g(\bm{r}_p),
        \label{eq:ms_phase_profile}
\end{equation}
for any meta-atom $p$ and any frequency $\omega$. Equation~\eqref{eq:ms_phase_profile} is the \textit{achromaticity condition}. It states that the phase profile of the metasurfaces is the summation of two terms, (i) the target phase profile~\eqref{eq:target_phase_profile} evaluated at the positions of each meta-atom $\bm{r}_p$ and (ii) a function $\tilde{\varphi}_0(\omega)$ of the frequency but independent of the position. The choice of the reference meta-atom appears to be a degree of freedom in the design of achromatic metasurfaces. In the literature, the term $\tilde{\varphi}_0$ is often introduced as $C(\omega)$ \cite{aieta_multiwavelength_2015, shrestha_broadband_2018}.
The design of achromatic metasurfaces usually involves
calculating the optical response, specifically the transmission
amplitude and phase, of an ensemble of meta-atoms
as detailed later in Section \ref{sec:dataset} \cite{shrestha_broadband_2018}. From this set of responses,
we must identify the subset of meta-atoms that
satisfy equation \eqref{eq:ms_phase_profile} at every position on the metasurface.
We refer to this selected group as the \textit{achromatic subset},
denoted as  $\mathcal{D}(\tilde{\varphi}_0)$ in the following discussion. The requirement of only selecting meta-atoms from $\mathcal{D}(\tilde{\varphi}_0)$ is analogous to the condition that the refractive index of conventional refractive lenses should be dispersion-less (see Suppl. Mat.~\ref{sm-sec:achrom_cond}). 

\subsubsection{Figure-of-Merit}
The optical performances of the metasurface can be described by the Strehl ratio, given by 
\begin{equation}
    S(\omega) =  \abs{\braket{e^{i\delta\varphi_p(\omega)}}}^2
    \;\text{with}\;
    \delta\varphi_p(\omega) = \varphi(\bm{r}_p, \omega) - \tilde{\varphi}_p(\omega)
    \label{eq:strehl_def}
\end{equation}
In Eq.~\eqref{eq:strehl_def}, the braket notation $\braket{\cdot}$ represents the spatial average over the selected meta-atoms. This quantity can be seen as a measure of the error on the metasurface phase profile.  To quantify the overall performances of the metasurfaces over the bandwidth, we introduce the figure-of-merit $\bar{S}$ as the mean of the Strehl ratio over the bandwidth $\Delta\omega$,
\begin{equation}
    \bar{S} = \frac{1}{\Delta\omega} \int_{\Delta\omega} S(\omega) d\omega.
    \label{eq:strehl_mean}
\end{equation}
A metasurface made of meta-atoms, whose at every position $\bm{r}_p$, satisfy the achromaticity conditions~\eqref{eq:ms_phase_profile}, has a unitary mean Strehl ratio, $\bar{S} = 1$, and it is said to be \textit{perfectly} achromatic as it produces an aberration-free wavefront over bandwidth, as represented in Fig.~\ref{fig:figure_0} (a). We emphasize that our figure-of-merit $\bar{S}$ quantifies both chromatic aberrations and geometrical aberrations since, from a design perspective, both aberrations should be minimized concurrently. Strictly, one would have to look at the variance of the Strehl ratio over the bandwidth to assess the chromatic aberrations independently of the geometric aberrations.

\subsubsection{Fundamental Bounds}
Having established the dispersion requirements for the meta-atoms, we now examine how the properties of the achromatic subset constrains the design of a perfectly achromatic metasurface. Specifically, the phase delay (or group delay) of meta-atoms belonging to an achromatic subset $\mathcal{D}(\tilde{\varphi}_0)$ is necessarily bounded, which must constrain the achievable size and optical functionalities. This raises the two fundamental questions: What is the largest metasurface that can be realized ? And how does this maximum size, $D$, scale with the desired operational bandwidth, $\Delta \omega$ ?

To answer these questions, we consider the case of a transmissive metasurface. The difference of phase delay between two plane waves, with frequencies $\omega_1$ and $\omega_2$, propagating through two different meta-atoms $p$ and $q$ of an achromatic subset $\mathcal{D}(\tilde{\varphi}_0)$ is necessarily bounded by a finite quantity $\Delta \tau$,
\begin{equation}
    \abs{\frac{\tilde{\varphi}_q(\omega_2)}{\omega_2} - \frac{\tilde{\varphi}_p(\omega_1)}{\omega_1}} \leq \Delta \tau,
    \label{eq:diff_np}
\end{equation}
Because the meta-atoms are part of the same achromatic subset $\mathcal{D}(\tilde{\varphi}_0)$, we can apply the achromaticity condition~\eqref{eq:ms_phase_profile}. It leads to, see Suppl. Mat.~\ref{sm-sec:size_cond},
\begin{equation}
    \frac{\Delta g}{c} + \Delta \tau_0 \leq \Delta \tau.
    \label{eq:size_cond}
\end{equation}
where $\frac{\Delta g}{c}$ is the maximal delay induced by the variation of optical path difference $\Delta g$ of the optical function; $\Delta \tau_0 = \max_{\omega_1,\omega_2} \abs{\frac{\tilde{\varphi}_0(\omega_2)}{\omega_2} - \frac{\tilde{\varphi}_0(\omega_1)}{\omega_1}}$ quantifies the maximal phase delay associated with the dispersion of the reference meta-atom. Larger values of $\Delta\tau_0$ indicate higher dispersive behavior of the meta-atoms, while a null value means the reference meta-atom is dispersion-less, \textit{i.e.} plane waves with different frequencies exit the metasurface at the same time. As a first approximation, for the case of effective refractive index metasurfaces, we can assume $\Delta\tau$ is bounded by a quantity proportional to the constitutive materials of the metasurfaces, $\Delta\tau \approx (n_{mat} - 1) \frac{h}{c}$.

Equation~\eqref{eq:size_cond} describes the trade-off between (i) the total variation of optical path difference of the target optical function and (ii) the dispersion of the selected achromatic subset. If the meta-atoms of the achromatic subset are dispersion-less, $\Delta\tau_0 = 0$, then the difference of optical path achievable is maximal $\Delta g \leq c \Delta \tau$,  meaning it is solely constrained by the bound $\Delta \tau$. This result shows that dispersion impacts negatively the achievable optical function when realizing achromatic metasurfaces. Translating this results in terms of size of the metasurface, $D \leq g^{-1}(c\Delta \tau - \Delta\tau_0)$, with dispersion-less meta-atoms we can achieve the largest achromatic metasurfaces. The maximum achievable size of the metasurface will decrease for increasing dispersion $\Delta\tau_0$ of the meta-atoms. This comes in contrast with approaches inspired by slow light which maximize the dispersion in order achieve larger scale achromatic metasurfaces. This will be further discussed below.


For an achromatic subset, the largest allowed phase delay is $\Delta\tau = \max_{(p,q),(\omega_1,\omega_2)}\abs{\frac{\tilde{\varphi}_q(\omega_2)}{\omega_2} - \frac{\tilde{\varphi}_p(\omega_1)}{\omega_1}}$. A visual representation of this relationship is provided in Fig.~\ref{fig:figure_0} (b).

For practical and numerical applications, when working with \textit{effective refractive index} metasurfaces made of high aspect ratio dielectric pillars, it is convenient to express the time delays in terms of effective refractive indices $n_p = \frac{c}{\omega h} \tilde{\varphi}_p$ of the meta-atoms (where $h$ is the height of the meta-atoms). For an achromatic metalens with given numerical aperture (NA) and achromatic subset $\mathcal{D}(\tilde{\varphi}_0)$, the maximal achievable size is
\begin{equation}
    D = 2 h (\Delta n - \Delta n_0) \frac{\text{NA}}{1 - \sqrt{1 - \text{NA}^2}},
    \label{eq:size_na}
\end{equation}
where $\Delta n$ is the maximal variation of effective refractive index over the bandwidth and achromatic subset, and $\Delta n_0$ is the variation of effective refractive index of the reference meta-atom over the bandwidth. Fig.~\ref{fig:figure_0} (c) illustrates  Eq.~\eqref{eq:size_na}, showing the normalized metasurface diameter $\frac{D}{h\Delta n}$, as a function of the numerical aperture $\text{NA}$ and the relative dispersion $\Delta n_0 / \Delta n$ for an achromatic meta-atom dataset. For a representative dataset with $h = \SI{1}{\micro\metre}$ and $\Delta n = 1$, a numerical aperture of $0.1$ and a relative dispersion of $10\%$ ($\Delta n_0 = 0.1$), imposing unitary Strehl ratio restricts the metalens diameter to below \SI{50}{\micro\metre}. Figure~\ref{fig:figure_0} (c) shows that, even under relatively modest constraints, the maximal achievable size of a perfectly achromatic metasurface remains quite limited. Consequently, achieving large-area metasurfaces with perfectly achromatic performance is expected to be highly challenging.

Note that Eq.~\eqref{eq:size_na} does not explicitly depend on the bandwidth $\Delta\omega$. It only appears implicitly through the expressions of the variation of effective refractive indices $\Delta n$ and $\Delta n_0$. But, when meta-atoms belong to the same achromatic subset, they share the same dispersion properties, which causes the term $\Delta n -\Delta n_0$ to become independent of the bandwidth, see. Suppl. Mat.~\ref{sm-sec:bandwidth_achromatic}. For a fixed achromatic subset $\mathcal{D}(\tilde{\varphi}_0)$, the maximal achievable size of a perfectly achromatic metasurface is independent of its operational bandwidth.

\subsection{Quasi-Achromatic Metasurfaces}\label{sec:quasi_achromaticity}
So far, we focused on perfectly achromatic metasurfaces, composed of meta-atoms belonging to the same achromatic subset $\mathcal{D}(\tilde{\varphi}_0)$. We demonstrated that such metasurfaces have optimal performances, $\bar{S} = 1$, but the optical function, characterized by the largest variation of optical path difference, is strongly constrained by the dispersion of the meta-atoms associated with the achromatic subset.

In practice, perfect performances may not always be necessary. Many optical systems can tolerate a certain degree of aberrations. For example, a Strehl ratio of $0.8$ is commonly regarded as sufficient for imaging applications. This raises an important question: How much can we relax the constraints on the optical function if we allow the metasurface to have a non-unitary Strehl ratio, and thus, use meta-atoms that do not belong to a single achromatic subset ?

To derive a trade-off condition linking optical performance, bandwidth, and size of the metasurfaces, we analyze the spectral phase of the meta-atoms in terms of its frequency derivatives (group delay and higher-order dispersions) as often presented in the literature \cite{chen_broadband_2018}. We first reinterpret the achromaticity conditions from this perspective and define the chromatic aberrations of the wavefront. Then, we perform an expansion of the Strehl ratio $\bar{S}$ in terms of chromatic aberrations in order to obtain a trade-off condition.

\begin{figure*}
    \centering
    \includegraphics[width=\textwidth]{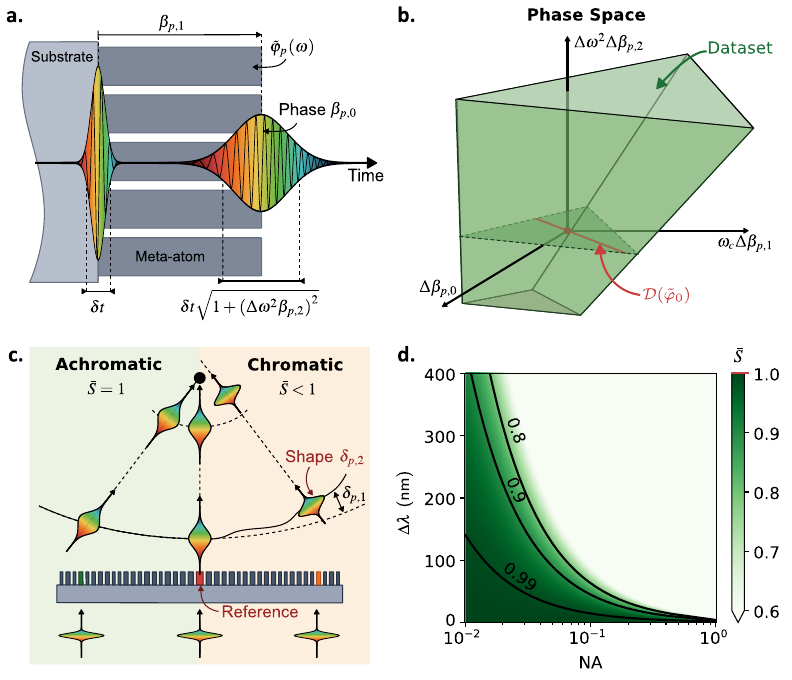}
    \caption[Quasi-achromatism in Metasurfaces]{\textbf{Quasi-achromatism in metasurfaces.} \textbf{a.} Schematic representation of the propagation of a Gaussian pulse through a metasurface with phase dispersion $\tilde{\varphi}_p$, described by Eq.~\eqref{eq:taylor_exp}. The pulse, initially with temporal width $\delta t$, exits the metasurface with time corresponding to the group delay $\beta_{p,1}$ and broadens with FWHM expressed in terms of group delay dispersion $\beta_{p,2}$. The phase of the oscillations inside the pulse is given by $\beta_{p,0}$. \textbf{b.} Schematic representation of the dataset, green volume, in phase space. The achromatic subset associated with the reference meta-atom, $\mathcal{D}(\tilde{\varphi}_0)$ is represented with a red line, as the intersection of the green volume with the line of equations $\Delta\beta_{p,0}=\omega_c \Delta\beta_{p,1}$ and $\Delta\omega^2\Delta\beta_{p,2}=0$. \textbf{c.} Schematic representation of the chromatic aberrations of the wavefront for a converging lens. In the achromatic regime, all the pulses exit the metasurfaces with the target delay and the same shape as the reference placed at the center. In the presence of aberrations, pulses have different shapes and exit the metasurface at different times. They do not interfere constructively at the focal point. \textbf{d.} Evolution of the average Strehl ratio $\bar{S}$ as function of the numerical aperture NA and the bandwidth $\Delta\lambda$. The iso-Strehl contours have been plotted in black plain lines with $\bar{S} = 0.99$, $0.9$ and $0.8$ respectively, and the parameter is chosen to be $\alpha = 0.1$.}
	\label{fig:figure_1}
\end{figure*}

\subsubsection{Chromatic Aberrations of the Wavefront}
Since we are no longer limited to an achromatic subset of meta-atoms, but instead we consider the full dataset, we must select the meta-atoms that minimize phase aberrations across the bandwidth to a specified degree. One approach to achieve this is by minimizing the uniform norm of the phase error over the entire bandwidth at each position of the metasurface, $\norm{\varphi - \tilde{\varphi}_p}_\infty$, thus approaching the achromaticity condition expressed in \eqref{eq:ms_phase_profile}. Alternative selection strategies are discussed in Suppl. Mat.~\ref{sm-sec:selection_strategies}. With this approach, the set of usable meta-atoms expands to include not only the achromatic subset $\mathcal{D}(\tilde{\varphi}_0)$ but also the set of sub-optimal meta-atoms introducing chromatic aberrations. However, for sufficiently small aberrations, we can consider the metasurface as \textit{quasi-achromatic}, meaning its mean Strehl ratio is close to unity.

To quantify the chromatic aberrations, we first assume that the spectral phase of each meta-atom varies smoothly over the operating bandwidth so that we can approximate the phase dispersion of the meta-atoms by a Taylor expansion, at the central frequency, over the bandwidth $\Delta\omega$,
\begin{equation}
    \tilde{\varphi}_p(\omega) = \beta_{p,0} + (\omega-\omega_c) \beta_{p,1} + \frac{1}{2}(\omega - \omega_c)^2 \beta_{p,2},
    \label{eq:taylor_exp}
\end{equation}
where $\beta_{p,0} = \tilde{\varphi}_p(\omega_c)$ is the phase shift at central frequency, $\beta_{p,1} = \left.\frac{d\tilde{\varphi}_p}{d\omega}\right\rvert_{\omega_c}$ the group delay and $\beta_{p,2} = \left.\frac{d^2\tilde{\varphi}_p}{d\omega^2}\right\rvert_{\omega_c}$ the group delay dispersion. Anticipating our numerical results, we can consider approximation up to the second order over the visible bandwidth. A visual illustration of how those quantities affect the propagation of a Gaussian pulse through the metasurface is provided in Fig.~\ref{fig:figure_1}(a).

By differentiating Eq.~\eqref{eq:ms_phase_profile} twice and evaluating at $\omega_c$, we obtain another expression of the \textit{achromaticity condition}. It forms a system of equations that must be satisfied simultaneously, for any meta-atom $p$
\begin{subequations}
    \begin{empheq}[left=\empheqlbrace]{align}
	\label{eq:achromaticity_cond_a}
	&\Delta\beta_{p,0} = \frac{\omega_c}{c}g(\bm{r}_p),\\
	\label{eq:achromaticity_cond_b}
	&\Delta\beta_{p,1} = \frac{1}{c}g(\bm{r}_p),\\
	\label{eq:achromaticity_cond_c}
	&\Delta\beta_{p,2} = 0,
     \end{empheq}
     \label{eq:achromaticity_3conds}
\end{subequations}
with $\Delta\beta_{p,m} = \beta_{p,m} - \beta_{0,m}$, for $m = 0,1,2$.  The achromaticity conditions~\eqref{eq:achromaticity_cond_a}-\eqref{eq:achromaticity_cond_c} are analogous to those proposed in \cite{chen_broadband_2018}. Condition~\eqref{eq:achromaticity_cond_a} enforces the phase profile at the central frequency. Equation~\eqref{eq:achromaticity_cond_b} ensures that Gaussian pulses emerge with the target time delay. Finally, Eq.~\eqref{eq:achromaticity_cond_c} requires that Gaussian pulses should have the same envelope, in the temporal domain, across the metasurface. Under the assumption in Eq.~\eqref{eq:taylor_exp}, Eq.~\eqref{eq:ms_phase_profile} and the system~\eqref{eq:achromaticity_cond_a}-\eqref{eq:achromaticity_cond_c} are equivalent. Specifically, the quantity $\Delta\beta_{\Max,1} = \max_{p\in\mathcal{D}(\tilde{\varphi}_0)} \abs{\Delta\beta_{p,1}}$ which quantifies the range of group delays spanned by an achromatic subset  can be shown to be equal to $\Delta\tau-\Delta\tau_0$ used in Eq.~\eqref{eq:size_cond}, see Suppl. Mat.~\ref{sm-sec:bandwidth_achromatic}. 

The phase dispersion of the meta-atoms from a dataset can be uniquely identified by the triplet $\beta_{p,0}$, $\beta_{p,1}$ and $\beta_{p,2}$. Each meta-atom can therefore be represented as a point in a three-dimensional \textit{phase space} with coordinates $(\Delta\beta_{p,0},\omega_c\Delta\beta_{p,1},\Delta\omega^2\Delta\beta_{p,2})$ and where the reference meta-atom is placed at the origin. Since these quantities are bounded, all meta-atoms of the dataset are expected to be confined within a volume, as represented in Fig.~\ref{fig:figure_1}(b). From the achromaticity condition~\eqref{eq:achromaticity_cond_a}-\eqref{eq:achromaticity_cond_c}, we can precisely interpret the achromatic subset $\mathcal{D}(\tilde{\varphi}_0)$, in phase space, as the intersection between constant group delay dispersion planes, \textit{i.e.} $\Delta\omega^2\Delta\beta_{p,2}=0$ and the plane with equation $\Delta\beta_{p,0}=\omega_c \Delta\beta_{p,1}$. It is visualized in Fig.~\ref{fig:figure_1}(b) by the red line.
 
Meta-atoms failing to simultaneously satisfy Eq.~\eqref{eq:achromaticity_cond_a}-\eqref{eq:achromaticity_cond_c} introduce chromatic aberrations. We denote by $\delta_{p,m}$ the $m$-th-order chromatic aberrations on the wavefront associated with the $p$-th meta-atoms placed at $\bm{r}_p$. These aberrations are given by
\begin{subequations}
    \begin{empheq}[left=\empheqlbrace]{align}
	\label{eq:wavefront_aberration_a}
	&\delta_{p,0} = \frac{\omega_c}{c}g(\bm{r}_p) - \Delta\beta_{p,0},\\
	\label{eq:wavefront_aberration_b}
	&\delta_{p,1} = \frac{1}{c}g(\bm{r}_p) - \Delta\beta_{p,1},\\
	\label{eq:wavefront_aberration_c}
	&\delta_{p,2} = - \Delta\beta_{p,2},
     \end{empheq}
     \label{eq:wavefront_aberration}
\end{subequations}
Equation~\eqref{eq:wavefront_aberration_a} quantifies the $0$-th order chromatic aberration, \textit{i.e.} it corresponds to geometrical aberrations of the wavefront at the central frequency. Equations~\eqref{eq:wavefront_aberration_b} and~\eqref{eq:wavefront_aberration_b} quantifies the $1$-st and $2$-nd order chromatic aberrations. These correspond, respectively, to the errors in the time at which the pulse exits the metasurface and in the expected shape of the pulse. One can interpret the $1$-st order chromatic aberrations by analogy of the geometrical aberrations for a pulse instead of a plane wave. These aberrations are represented in Fig.~\ref{fig:figure_1}(c).

As a final remark on the chromatic aberrations, it should be noted that considering only the second-order Taylor expansion implicitly limits the valid bandwidth for the metasurface. For very large bandwidths, higher-order group delay dispersion terms must be taken into account. This would require adding subsequent equations to the system~\eqref{eq:achromaticity_cond_a}-\eqref{eq:achromaticity_cond_c}, thereby taking into consideration higher order chromatic aberrations. As a result, finding meta-atoms that satisfy all conditions simultaneously becomes increasingly challenging. 

\subsubsection{Mean Strehl Ratio Expansion}
Having defined the chromatic aberrations, we can now evaluate their impact on the optical performances, specifically the average Strehl ratio $\bar{S}$ introduced previously, in the case of quasi-achromaticity. Assuming relatively small aberrations, we can expand the Strehl ratio \eqref{eq:strehl_def} at each frequency, in terms of wavefront aberrations, cf. Suppl. Mat.~\ref{sm-sec:strehl_ratio},
\begin{align}
    \nonumber
    S(\omega) \approx 1 - V_{0,0} &- 2(\omega-\omega_c) V_{0,1} - (\omega-\omega_c)^2 \left( V_{1,1} + V_{0,2} \right)\\
    & - (\omega-\omega_c)^3 V_{1,2} - \frac{(\omega-\omega_c)^4}{4} V_{2,2},
    \label{eq:strehl_expansion}
\end{align}
with $V_{m,n}$ the covariance between the chromatic aberrations of the wavefront of order $m$ and $n$,
\begin{equation}
    V_{m,n} = \braket{\delta_m\delta_n} - \braket{\delta_m}\braket{\delta_n}.
    \label{eq:V_pq}
\end{equation}
The terms $V_{m,m}$ are the variances of the different orders of chromatic aberrations. Since these terms are always positive, they contribute to a reduction of the Strehl ratio. In contrast, the terms $V_{m,n}$, with $m\neq n$, represent the covariances which quantify the correlations among different orders of aberration. These covariances can be positive or negative, allowing partial compensation effects. Essentially, we can tolerate slight variations in the shape of pulses exiting the metasurfaces if these variations are properly compensated by adjusting the pulse delay, thereby limiting the average deterioration of the Strehl ratio.

Our figure-of-merit $\bar{S}$ can then be deduced from Eq.~\eqref{eq:strehl_expansion} by taking the mean over the bandwidth, leading to
\begin{equation}
    \bar{S} = 1 - V_{0,0} - \frac{V_{1,1} + V_{0,2}}{12} \Delta\omega^2 - \frac{V_{2,2}}{320} \Delta\omega^4.
    \label{eq:strehl_mean_exp}
\end{equation}
The average Strehl ratio evolves in $\Delta\omega^2$ and $\Delta\omega^4$ for the first and second orders chromatic aberrations, respectively. This means that the impact of chromatic aberrations becomes increasingly significant as the operational bandwidth increases. If the metasurface is composed exclusively of meta-atoms exhibiting neither $1$-st nor $2$-nd order chromatic aberrations, the Strehl ratio is non-unitary, yet the metasurface remains achromatic. In this sense, the $0$-th order quantifies the geometric aberrations of the metasurface over the bandwidth.

\subsubsection{Trade-off Condition}
Equation~\eqref{eq:strehl_mean_exp} relates the optical performances to the variance of the chromatic aberrations. In principle, it is difficult to anticipate what these aberrations will be, and so, to estimate their variances. To highlight the relationship between optical performance $\bar{S}$, size of the metasurface $D$ and bandwidth $\Delta\omega$, we can use Eq.~\eqref{eq:strehl_mean_exp} in simplified cases where we parametrize the variance of the chromatic aberrations.

For simplicity, we consider the scenario in which the meta-atoms exhibit only $1$-st order chromatic aberrations, \textit{i.e.} $V_{1,1} \neq 0$. We neglect the $2$-nd order chromatic aberrations and consider $V_{0,0} = 0$ since it quantifies only geometric aberrations and corresponds to a general offset of the mean Strehl ratio. In this case, the metasurface is characterized by four parameters $(D, \Delta\omega, \bar{S}, V_{1,1})$. It is important to emphasize that the optical performance of the metasurface depends only on the variance of the chromatic aberrations and not on their spatial distribution across the metasurface. In other words, two metasurfaces with the same size $D$, operating bandwidth $\Delta\omega$ and aberration variance $V_{1,1}$, will share the same average Strehl ratio $\bar{S}$, even if the aberrations are distributed differently among their constituent meta-atoms.

We parametrize the variance $V_{1,1}$ of the $1$-st order chromatic aberrations such that the standard deviation $\sqrt{V_{1,1}}$ represents a percentage $\alpha$ of error on the group delay associated with the outermost meta-atoms, \textit{i.e.} $\sqrt{V_{1,1}} = \alpha \frac{g(D)}{c}$. For the particular case of a metalens with diameter $D$ and a numerical aperture $\text{NA}$, this relation leads to the following expression (see Suppl. Mat.~\ref{sm-sec:mean_strehl})
\begin{equation}
    \frac{\Delta\omega D}{\sqrt{1 - \bar{S}}} = \frac{4\sqrt{3} c}{\alpha} \frac{\text{NA}}{1 - \sqrt{1 - \text{NA}^2}},
    \label{eq:strehl_bandwidth}
\end{equation}
where $\alpha \neq 0$, and $\bar{S}\neq 1$ by definition. Due to our initial assumptions, this equation is valid only in the quasi-achromatic regime, with near unity $\bar{S}$. 

Equation~\eqref{eq:strehl_bandwidth} provides a qualitative description of the size-bandwidth-Strehl ratio trade-off for a quasi-achromatic metasurface with a fixed standard deviation of the $1$-st order chromatic aberrations. For a given numerical aperture NA and non-unitary Strehl ratio, we obtain an inverse law of the size $D$ as function of the bandwidth, demonstrating that in the quasi-achromatic regime, the size of the metasurface depends on the bandwidth. Figure.~\ref{fig:figure_1}(d) illustrates this trade-off condition as function of the numerical aperture and the bandwidth. It shows that in this already favorable scenario, with a moderate amount of aberrations $\alpha = 0.1$ and Strehl ratio $\bar{S} = 0.8$, one may not expect to realize a numerical aperture larger of $0.1$ for a bandwidth larger than \SI{100}{\nano\metre}.

\section{Numerical Results}
\subsection{Dataset Characterization}
\label{sec:dataset}

\begin{figure*}
    \centering
    \includegraphics[width=\textwidth]{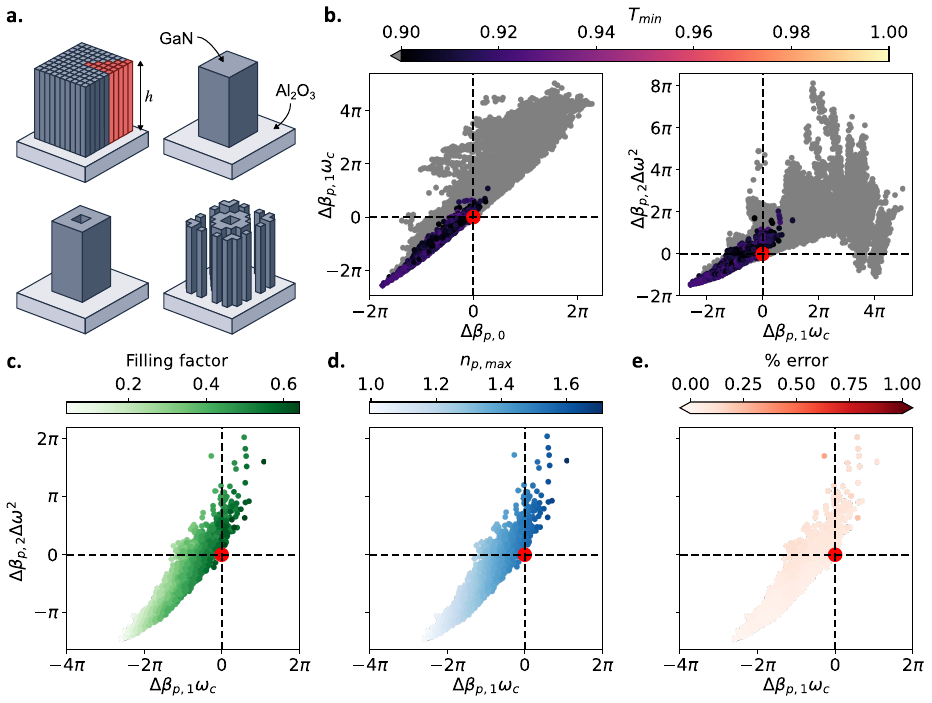}
    \caption[Characterization of the dataset]{\textbf{Characterization of the dataset}. \textbf{a.} Schematic representation of the \ce{GaN} pillars with height $h$ which have been simulated. The irreducible triangle is represented in red. \textbf{b.} Representation of the projection of the whole dataset in the planes ($\Delta\beta_{p,1}\omega_c$,$\Delta\beta_{p,2}\Delta\omega^2$) and ($\Delta\beta_{p,0}$,$\Delta\beta_{p,1}\omega_c$) of the phase space. \textbf{c.} Representation of the filling factor of \ce{GaN} in the unit-cell of each meta-atoms of the curated dataset. \textbf{d.} Representation of the maximum of the effective refractive index, over the bandwidth, of each meta-atoms. \textbf{e.} Representation of the relative percentage error $\varepsilon=\frac{\left\lVert \tilde{\varphi}_p-\varphi_{\text{Taylor}} \right\rVert_\infty}{\left\lVert \tilde{\varphi}_p \right \rVert_\infty}$ made on the phase dispersion $\tilde{\varphi}_p$ when approximating its second degree Taylor expansion at central frequency. The phases $\phi_{\text{Taylor}}$ are calculated by \eqref{eq:taylor_exp} and the infinite norm corresponds to the maximum error over the bandwidth.}
    \label{fig:figure_2}
\end{figure*}

To validate our theory, we built a dataset of meta-atoms made of gallium nitride (\ce{GaN}) pillars, with height $h=\SI{1}{\micro\metre}$, on a sapphire (\ce{Al_2O_3}) substrate. The period of the metasurface is $p=\SI{350}{\nano\metre}$, ensuring to be in the metasurface regime for the visible range.

The geometry of the meta-atoms has been defined pixel by pixel following the methodology in \cite{shrestha_broadband_2018} (additional materials). A meta-atom is composed of a $10\times 10$ array of pixels with size $\SI{30}{\nano\metre}\times\SI{30}{\nano\metre}$ each. Every pixel can be filled with air or with \ce{GaN}. To obtain polarization-insensitive structures, the meta-atoms have to be $C_4$-symmetric. We thus only control the $15$ pixels of the irreducible triangle, as depicted Fig.~\ref{fig:figure_2} (a), and generate the rest by symmetry. Consequently, we obtain a dataset of $2^{15}=32768$ different meta-atoms, with a wide variety of geometries including squared pillars (top right), hollowed structures (bottom left), but excluding rotationally symmetric structures.

The meta-atoms have been simulated by finite-domain time-difference (FDTD) simulations, with commercial software Lumerical, following the traditional \textit{lookup} table approach \cite{chen_flat_2020}. We extract the phase dispersions and transmission dispersions of the transmitted electric field in the visible range, $400-\SI{800}{\nano\metre}$, for each meta-atom of the dataset. We compute then the corresponding coefficients $\Delta\beta_{p,i}, i=0,1,2$, relative to a given reference meta-atom. The results are summarized in Fig.~\ref{fig:figure_2}, where each meta-atom dispersion is represented as a point.

In Fig.~\ref{fig:figure_2} (b), we represent the projection of the whole dataset in the planes $(\Delta\beta_{p,0}, \omega_c\Delta\beta_{p,1})$ and $(\omega_c\Delta\beta_{p,1}, \Delta\omega^2\Delta\beta_{p,2})$ of the phase space. Since we are interested in high transmission meta-atoms, we exclude those with a minimum transmission lower that $0.9$, which are represented in gray. The remaining dataset, comprising approximately 4000 meta-atoms, is displayed using the purple-orange color scale. The shape of the dataset reveals a clear correlation between phase, group delay and group delay dispersion. Meta-atoms with smaller phase delay are associated with smaller group delay and group delay dispersion, which is consistent with \cite{shim_fundamental_2021}.
From Eq.~\eqref{eq:size_cond}, it is clear that one needs to maximize $\Delta\tau-\Delta\tau_0$ in order to realize large achromatic metasurfaces. Given that $\Delta\tau-\Delta\tau_0$ is equal to the range spanned by $\Delta\beta_{p,1}$, this amounts to selecting the achromatic subset spanning range of $\Delta\beta_{p,1}$ in Fig.~\ref{fig:figure_2}.
In other words, in Eq.~\eqref{eq:size_cond}, the available range of the maximal phase delay difference $\Delta\tau$ of an achromatic subset is not independent of $\Delta\tau_0$, meaning less dispersive achromatic subsets can only induce small variations of optical path difference. This is a first compromise limiting optical functionality of achromatic metasurfaces. In order to realize a converging lens, we choose the origin of the phase space, \textit{i.e.} the reference meta-atom in red, on the top-right side of the dataset in order to maximize the size of the achromatic subset.

To give a visual representation of how the geometries affect the dispersion of the meta-atoms, we represented in Fig.~\ref{fig:figure_2} (c) the \ce{GaN} filling factor of the curated dataset in the projection plane $(\omega_c\Delta\beta_{p,1}, \Delta\omega^2\Delta\beta_{p,2})$. It shows that meta-atoms with a larger fraction of \ce{GaN} are associated to larger group delays and group delay dispersions (top-right), while the meta-atoms with the smallest fraction are associated with the lower group delay and group delay dispersions (bottom-left). Some meta-atoms sharing the same filling factor exhibit different group delay $\Delta\beta_{p,1}$ and group delay dispersion $\Delta\beta_{p,2}$, suggesting that the geometry plays a role in the dispersion. However, this effect is relatively small in our dataset, which we attribute to the pixelization of the meta-atoms, thus excluding cylindrical geometries \cite{shrestha_broadband_2018}.

Since those meta-atoms are high-aspect ratio pillars, we determined their effective refractive indices $n_p$ and plotted their maximal value over the bandwidth in Fig.~\ref{fig:figure_2} (d). We notice that the maximum effective refractive index of the curated dataset does not exceed a value close to $n_{p,\Max} \approx 1.7$ well below the refractive index of \ce{GaN} ($n_{\ce{GaN}} \approx 2.5$). This gap can likely be attributed to the exclusion of meta-atoms that support resonances. In simple terms, a resonance in a metasurface arises when light can be diffracted into quasi-guided waves that propagate laterally within the metasurface plane \cite{kazarinov1985second}. For such quasi-guided waves to exist, a necessary (though not sufficient) condition is that the effective refractive index of the grating must surpass that of the surrounding medium. Consequently, resonances begin to appear only when the effective refractive index exceeds that of the \ce{Al_2O_3} substrate ($n_{\ce{Al_2O_3}} = 1.77$). By excluding meta-atoms inducing resonances, the available range of effective refractive indices is further limited to values between $1$ and the substrate. 

Finally, Fig.~\ref{fig:figure_2} (e) shows the relative maximum phase difference across the bandwidth between the real phase and the Taylor expansion approximation of this phase. Although the Taylor expansion is supposed to be a local approximation, the error stays below $1\%$ for all meta-atoms across the whole bandwidth, which justifies the use of the Taylor expansion for the subsequent analysis of the aberrations on the phase profile to a very good approximation.

\subsection{Theory Validation}\label{sec:validation}

\begin{figure*}
    \centering
    \includegraphics[width=\textwidth]{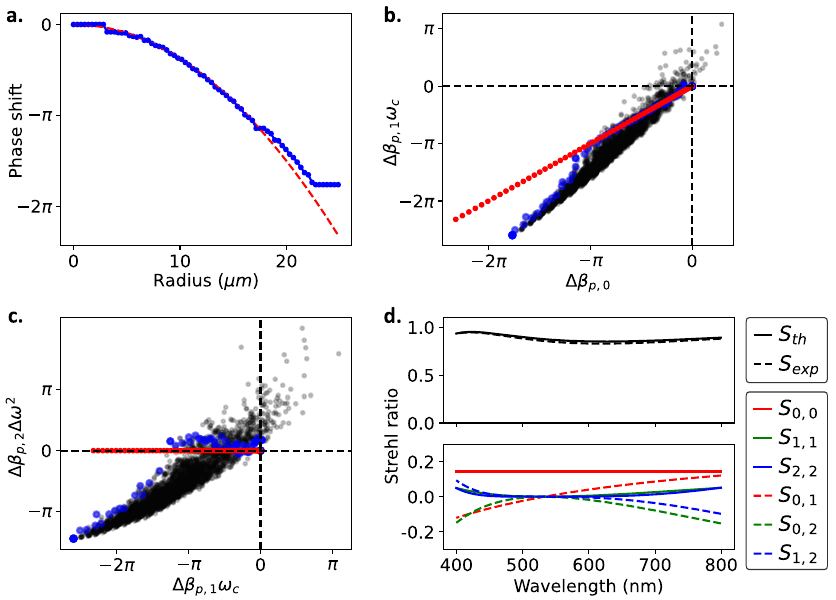}
    \caption[Achromatic metalens design]{\textbf{Achromatic metalens design.} \textbf{a.} Phase profile of the designed metalens at the central frequency $\omega_c$ in blue, and target phase profile at the same frequency in red dashed line. \textbf{b.} and \textbf{c.} Representation of the dataset in phase space in black and selected meta-atoms in blue. The red dots represent the target ideal meta-atoms that would form an achromatic subset. \textbf{d.} Strehl ratio as a function of the frequency. In the top panel, the black plain line represents the theoretical Strehl ratio while the dotted line represents the Strehl ratio computed using the expansion equation~\eqref{eq:strehl_expansion}. In the bottom panel, terms of the expansion are plotted, with in plain lines the terms corresponding to the variance of the chromatic wavefront aberrations and in dotted lines the terms associated with the covariance. The terms $S_{p,q}$ are given by $S_{0,0} = V_{0,0}$, $S_{1,1} = (\omega-\omega_c)^2V_{1,1}$, $S_{2,2} = \frac{1}{4}(\omega-\omega_c)^4 V_{2,2}$, $S_{0,1} = 2(\omega-\omega_c)V_{0,1}$, $S_{0,2} = (\omega-\omega_c)^2 V_{0,2}$ and $S_{1,2} = (\omega-\omega_c)^3 V_{1,2}$.}
    \label{fig:figure_3}
\end{figure*}

We use our dataset to realize an achromatic metalens with diameter $\SI{50}{\micro\metre}$ and focal length $\SI{500}{\micro\metre}$, cf. Fig.~\ref{fig:figure_3} (a). The optical path difference function $g(\bm{r})$ is given by the conventional equation $g(\bm{r}) = f - \sqrt{f^2 + \abs{\bm{r}}^2}$. The target phase profile of the metalens is plotted in phase space in Fig.~\ref{fig:figure_3} (b-c), red points, where we choose the reference meta-atom so that there is a significant overlap with the dataset of meta-atoms. These points are located on the line with equation $\Delta\beta_{p,0}=\omega_c\Delta\beta_{p,1}$, in Fig.~\ref{fig:figure_3} (b) and $\Delta\omega^2\Delta\beta_{p,2}=0$ in Fig.~\ref{fig:figure_3} (c). The selected meta-atoms, in blue, minimize the uniform norm $\norm{\tilde{\varphi}_p - \varphi}_\infty$ over the visible range. By definition of the reference meta-atom, the points closer to the origin of the phase space correspond to meta-atoms at the center of the metalens, while points at a larger distance correspond to meta-atoms closer to the border of the metalens. Their effects on the propagation of a Gaussian pulse can be readily determined by their location, in phase space, with respect to the reference meta-atom located at the center of the metalens: (i) $\Delta\beta_{p,0} \leq 0$, consistent with the expected behavior of a convergent monochromatic metalens in which the central meta-atom imparts a larger phase delay than peripheral elements; (ii) $\Delta\beta_{p,1} \leq 0$ and $\Delta\beta_{p,2} \leq 0$, indicating that meta-atoms situated at larger radial positions introduce reduced group delay and reduced temporal broadening relative to the reference, as it is also schematically represented in Fig.~\ref{fig:figure_1} (c). The subset of meta-atoms satisfying the achromatic conditions imposed by Eqs.~\eqref{eq:achromaticity_cond_a}-\eqref{eq:achromaticity_cond_c}, associated with the prescribed phase dispersion $\tilde{\varphi}_0$, corresponds to the set of blue points superposed to the red markers representing the target phase profile. On the contrary selected meta-atoms, in blue, deviating from the red markers correspond to meta-atoms introducing chromatic aberrations.  

The Strehl ratio $S(\omega)$ of the metalens has been computed at each frequency, using its exact expression~\eqref{eq:strehl_def} and compared with its expansion~\eqref{eq:strehl_expansion}, as shown in Fig.~\ref{fig:figure_3} (d). We observe good agreement between the exact value of the Strehl ratio and its expansion. Remarkably, the Strehl ratio is larger than $0.8$ over the bandwidth and reaches $0.9$ at shorter wavelengths. Looking at the expansion coefficients gives us insight into this behavior. The largest error occurs on the phase profile at the central frequency, corresponding to the $0$-th order chromatic aberration, with $S_{0,0} \approx 0.18$. This limitation is evidenced in Fig.~\ref{fig:figure_3} (b), where the required phase profile at the central frequency exceeds $2\pi$ at the metalens periphery, yet the available meta-atom library does not contain elements providing phase shifts beyond this value.

The target phase profile is defined such that the meta-atom dataset adequately spans the required group delay $\Delta\beta_{p,1}$, thereby limiting the residual error $S_{1,1} = (\omega-\omega_c)^2V_{1,1}$ associated with the first-order chromatic aberration. The error $S_{0,1}$ associated with the covariance terms $V_{0,1}$ are non-negligible; however, since this term is proportional to $(\omega - \omega_c)$, it changes sign at $\omega = \omega_c$. For frequencies $\omega < \omega_c$, one has $S_{0,1} > 0$, which implies $V_{0,1} < 0$: meta-atoms exhibiting the largest error in $\delta_{p,0}$ simultaneously present the smallest error in $\delta_{p,1}$. Consequently, the zeroth-order chromatic aberration—namely the error on the phase profile at the central frequency—is partially compensated by the first-order chromatic aberration, \textit{i.e.} the error on the group delay. An analogous analysis of the covariance terms $S_{0,2}$ yields $V_{0,2} < 0$.

\begin{figure*}
    \centering
    \includegraphics[width=\textwidth]{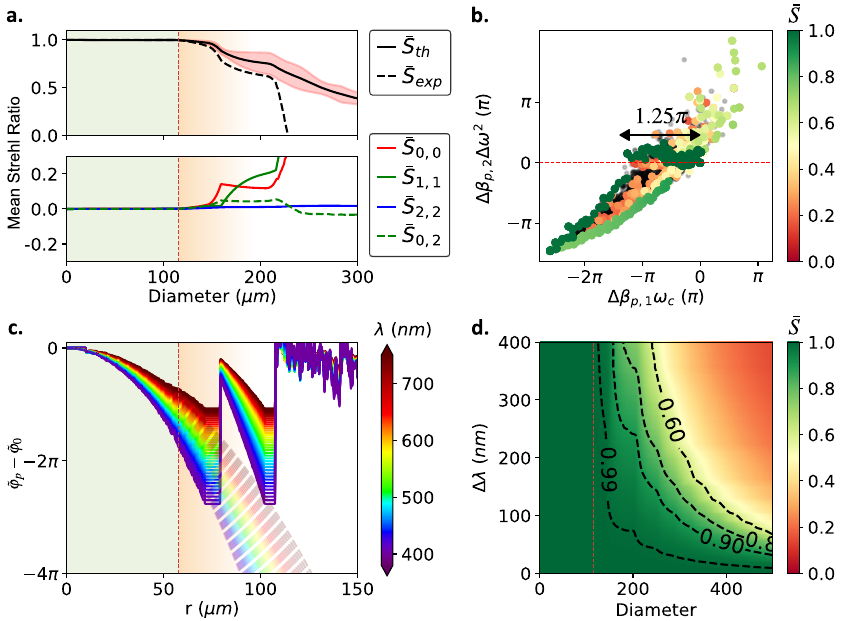}
    \caption[Achromatic and Chromatic Regimes in Metasurfaces]{\textbf{Achromatism in metasurfaces.} \textbf{a.} Evolution of the Strehl ratio as function of the diameter of a metalens with focal length $f = \SI{5}{\milli\metre}$. (Top panel) The black plain curve represents the exact Strehl ratio and the dashed line its expansion approximation. The red area represents the standard deviation of the Strehl ratio $S(\omega)$. (Bottom panel) Strehl ratio expansion coefficient with $\bar{S}_{0,0} = V_{0,0}$, $\bar{S}_{1,1} = \frac{\Delta\omega^2}{12}V_{1,1}$, $\bar{S}_{0,2} = \frac{\Delta\omega^2}{12}V_{0,2}$, and $\bar{S}_{2,2} = \frac{\Delta\omega^4}{320}V_{2,2}$. \textbf{b.} Representation in phase space of the selected meta-atoms of the dataset with color code representing the average Strehl ratio $\bar{S}$, and the arrow shows $\Delta\beta_{\Max,1}\omega_c = 1.25\pi$. \textbf{c.} Evolution of the phase profile of the metasurface as function of the position $\bm{r}_p$ at each frequency. The plain color lines represent the realized phase profile, and the dashed lines represent the target phase profile. The color map represents each wavelength. \textbf{d.} Evolution of the Strehl ratio as a function of the diameter of the metalens and the bandwidth. The dashed lines represent the iso-Strehl ratio contours. In \textbf{a.}, \textbf{b.} and \textbf{c.} the red and dashed vertical line represents the diameter of the achromatic metalens computed using Eq.~\eqref{eq:diameter}.}
    \label{fig:figure_4}
\end{figure*}

In the previous example, the metalens operates at low numerical aperture, such that the target phase profile lies largely within the range covered by the meta-atom library, except the metalens periphery, where the phase requirement exceeds the available range and a single meta-atom is necessarily repeated, resulting in a locally constant phase profile. Extending the aperture by simply tiling the boundary meta-atom would severely degrade the Strehl ratio, as quantified by Eq.~\eqref{eq:strehl_expansion}, owing to the accumulation of large phase errors. This degradation can be mitigated during the meta-atom selection step by replacing the standard error metric with the uniform norm $\|e^{i\tilde{\varphi}_p} - e^{i\varphi}\|_\infty$, thereby directly optimizing the Strehl ratio.

Let's now consider a metalens with fixed focal length, $f = \SI{5}{\milli\metre}$ with varying diameter, and hence varying numerical aperture. For each aperture, the metasurface is optimized to achieve achromaticity over the visible spectral range. Since the focal length is held constant, the target phase profile remains invariant, and increasing the aperture strictly reduces to appending additional meta-atoms at the periphery of the previously designed metalens. For each metalens, the spectrally averaged Strehl ratio $\bar{S}$ and its expansion coefficients are computed and displayed in Fig.~\ref{fig:figure_4} (a). For diameters below \SI{115}{\micro\metre} (green region), $\bar{S} = 1$, indicating that all constituent meta-atoms belong to the same achromatic subset. The corresponding maximum aperture can be determined analytically from Eq.~\eqref{eq:achromaticity_cond_b}
\begin{align}
    \nonumber
    D = 2f&\sqrt{\left[ 1 + \frac{c}{f} \Delta\beta_{\Max,1} \right]^2-1}\\
    \quad&\text{with}\quad
    \Delta\beta_{\Max,1} = \max_{p\in\mathcal{D}(\tilde{\varphi}_0)} \abs{\Delta\beta_{p,1}},
    \label{eq:diameter}
\end{align}
and using the value $\Delta\beta_{\Max,1}\,\omega_c = 1.25\pi$, extracted from Fig.~\ref{fig:figure_4} (b). Let us point out that, since $\Delta\beta_{\Max,1} = \Delta\tau - \Delta\tau_0$, Eq.~\eqref{eq:diameter} is fully equivalent to Eq.~\eqref{eq:size_cond} in case of equality.

The corresponding phase profiles, shown in Fig.~\ref{fig:figure_4} (c), confirm close agreement between the target and designed phase profiles across all wavelengths within this aperture range, consistent with the unitary average Strehl ratio and the perfectly achromatic regime. The standard deviation of the Strehl ratio over the bandwidth, plotted in red, vanishes identically in this regime. For diameters exceeding \SI{115}{\micro\metre}, the peripheral meta-atoms no longer satisfy the achromatic conditions~\eqref{eq:achromaticity_cond_a}--\eqref{eq:achromaticity_cond_c}, as evidenced by non-zero coefficients $\bar{S}_{0,0} = V_{0,0}$ and $\bar{S}_{1,1} = \frac{\Delta\omega^2}{12} V_{1,1}$ in Fig.~\ref{fig:figure_4} (a). Although the group delay dispersions of these outer meta-atoms deviates from that of the reference, its impact remains limited, as $\bar{S}_{2,2}$ stays negligibly small. Consequently, the phase profiles exhibit increasing deviation from the target, and the standard deviation of the Strehl ratio increases, leading to quasi-achromaticity. Eventually, at a certain aperture size, around \SI{150}{\micro\metre}, since we are minimizing a difference of complex exponential $\|e^{i\tilde{\varphi}_p} - e^{i\varphi}\|_\infty$, a phase wrapping occurs. Incidentally, the $0$-th order chromatic aberration decreases, but the $1$-st order sharply increases because none of the meta-atoms of the dataset can match the target group delay, leading to a broadening of the standard deviation of the Strehl ratio. After the second phase wrapping zone, the discrepancy between the target phase profile and the available quantities from our dataset becomes too significant. As a result, our method of selecting meta-atoms by minimizing the uniform norm produce a random pattern, see Suppl. Mat.~\ref{sm-sec:norm_choice}. At this point, the mean Strehl ratio is already too low to be considered acceptable.

To synthesize, we can define two regimes of operation for an achromatic metasurface. The first regime is the \textit{achromatic regime}, where the meta-atoms are part of an achromatic subset and satisfy Eq.~\eqref{eq:achromaticity_cond_a}-\eqref{eq:achromaticity_cond_c}. In the perfectly achromatic regime, the average Strehl ratio is unity and the aperture is uniquely determined by the achromatic conditions. Beyond this regime, whenever the dataset is not sufficient, the metasurface becomes chromatic—the constituent meta-atoms no longer exactly satisfy the achromatic conditions, and both the optical performance and the achievable aperture become bandwidth-dependent. A second operational regime, termed the \textit{quasi-achromatic} regime, can be defined as one in which the Strehl ratio remains close to unity despite the partial violation of the achromatic conditions.

This behavior is illustrated in Fig.~\ref{fig:figure_4} (d), which shows the average Strehl ratio as a function of both bandwidth and diameter for the metalens under consideration. The iso-Strehl contours, plotted as black dashed lines, display a vertical asymptote at $D = \SI{115}{\micro\metre}$, consistent with the perfectly achromatic regime in which the maximum achievable aperture is independent of the bandwidth. As the Strehl ratio constraint is relaxed, the iso-contours transition to an inverse-law dependence on diameter, reflecting the trade-off between aperture and bandwidth in the quasi-achromatic regime. This relationship between $\bar{S}$, $D$, and $\Delta\omega$ is captured qualitatively by Eq.~\eqref{eq:strehl_bandwidth}, as detailed in Suppl.~Mat.~\ref{sm-sec:mean_strehl}.

\section{Conclusion}\label{sec:discussion}
We have presented a rigorous framework for the design of achromatic metasurfaces, centered on the distinction between two operational regimes: (i) the perfectly achromatic regime, where all meta-atoms exactly satisfy the achromaticity conditions, yielding a unitary Strehl ratio across the full operating bandwidth, and bandwidth-independent optical function; (ii) the quasi-achromatic regime, where residual chromatic aberrations are tolerated, giving rise to a fundamental trade-off between aperture, bandwidth, and Strehl ratio derived assuming a variance-based parametrization of the aberrations.

Our results show that the size limit delineating the achromatic and quasi-achromatic regimes is exactly determined by the range of group delay associated to a chosen achromatic subset. This comes in agreement with previous works which established the group delay as principal limitation to the realization of large size or large numerical aperture achromatic metasurfaces. However, our work demonstrate that obtaining dataset with a large range of group delay is necessary but not sufficient. In order to reduce chromatic aberrations of the wavefront, meta-atoms must share the same effective index dispersion. Failing to meet this requirement introduces chromatic aberrations of the wavefront at the origin of previously established delay-bandwidth limits derived for achromatic metasurfaces.

The strict constraints governing the fabrication of achromatic metalenses limit them to small components with low numerical apertures. Similar conclusions apply to other types of metasurfaces, such as deflectors. While these metasurfaces may still have niche applications, many use cases will require alternative solutions that are not subjected to such constraints. For example, in some cases it may be sufficient to realize polychromatic metasurfaces whose optical performances are optimal only for a discrete set of frequencies instead of a continuous range \cite{aieta_multiwavelength_2015,chen_theory_2023}. Another alternative is to design compound systems, similar to traditional achromatic doublets, where the chromatic aberration of one component can be compensated by those of another. This can be realized with doublet of metasurfaces \cite{feng2022rgb} but also with hybrid systems composed of refractive components along with metasurfaces \cite{sawant2021aberration}. Finally, another promising perspective is to co-design optical systems in order to correct the chromatic aberrations introduced by metalens-based optical systems at the post-processing stage \cite{froch2025beating}.

\begin{acknowledgments}
PG, NK, SK acknowledge financial support from the French National Research Agency ANR Project Millesime 2 (ANR-21-ASTR-0014). SK acknowledges financial support from the European Defense fund EDF project Mini-Bot (EDF-2021-OPEN-R-SME-2). RC was supported by the French government through the France 2030 investment plan managed by the National Research Agency (ANR), as part of the Initiative of Excellence Université Côte d’Azur (ANR-15-IDEX-01). RC and SK acknowledge the support from European Innovation Council (EIC Pathfinder Open,TwistedNano, 2021–101046424). AP acknowledges the support from the European Union’s Horizon Marie Sklodowska-Curie Actions (RISE project 101007896 CHARTIST)
\end{acknowledgments}

\section*{Disclosures}
The authors declare no conflicts of interest.

\section*{Data Availability}
The data that support the findings of this study are available from the corresponding author upon reasonable request.

\section*{Supplemental document}
See supplementary materials for supporting content.

\bibliography{ref}

\end{document}


\title{Supplementary Materials - Beyond Group Delay: Fundamental Limits of Achromatic Metasurfaces}

\author{Roman Buisine}
\thanks{Equal contribution}
\affiliation{Université Côte d'Azur, CNRS-CRHEA, Rue Bernard Gregory, Valbonne, 06650, France.}%

\author{Nicolas Kossowski}
\thanks{Equal contribution}
\email{nicolas.kossowski@crhea.cnrs.fr}
\affiliation{Université Côte d'Azur, CNRS-CRHEA, Rue Bernard Gregory, Valbonne, 06650, France.}%

\author{Adelin Patoux}
\affiliation{Université Côte d'Azur, CNRS-CRHEA, Rue Bernard Gregory, Valbonne, 06650, France.}%

\author{Patrice Genevet}
\affiliation{Université Côte d'Azur, CNRS-CRHEA, Rue Bernard Gregory, Valbonne, 06650, France.}%
\affiliation{Department of Physics, Colorado School of Mines, 1523 Illinois St., Golden, CO, 80401, USA. }%

\author{Rémi Colom}
\affiliation{Université Côte d'Azur, CNRS-CRHEA, Rue Bernard Gregory, Valbonne, 06650, France.}%

\author{Samira Khadir}
\email{samira.khadir@crhea.cnrs.fr}
\affiliation{Université Côte d'Azur, CNRS-CRHEA, Rue Bernard Gregory, Valbonne, 06650, France.}%

\begin{abstract}
This supplement in the format provided by the authors
and unedited. Further distribution of this work must maintain attribution to the author(s) and the
published article’s title, journal citation, and DOI.
\end{abstract}

\maketitle

\section{Achromaticity Condition}\label{sec:achrom_cond}
We consider a metasurface with target phase profile given by
\begin{equation}
    \varphi(\bm{r},\omega) = \frac{\omega}{c} g(\bm{r}),
    \label{eq:opd_phase}
\end{equation}
where $g(\bm{r})$ is the optical function to realize, and $\bm{r}$ the coordinate position on the plane of the metasurface. The target phase profile is the product of two functions, one function of $\omega$ and one function of the position $g(\bm{r})$, commonly called optical path difference. Metasurfaces with this form of target phase profile have their phase profile derived from ray optics as the difference of optical path between two rays. Consequently, the target phase profile $\varphi$ is defined such that, for all $\omega$, $\varphi(0,\omega) = 0$. 
For a metasurface, we define by $\bm{r}_p$ the positions of the meta-atoms of the metasurface. At position $\bm{r} = \bm{r}_p$, we want to find an element such that
\begin{equation} 
    \varphi(\bm{r}_p,\omega) = \delta \tilde{\varphi}_p(\omega),
    \label{eq:varphi_d}
\end{equation}
with $\delta\tilde{\varphi}_p(\omega)$ the phase shift of the $p$-th meta-atom. The notation $\delta$ emphasizes that this phase shift of the meta-atom is defined with regards to a \textit{reference} meta-atom. Noting $\tilde{\varphi}_p(\omega)$, the \textit{phase dispersion} of the $p$-th meta-atom and choosing the reference to be $\tilde{\varphi}_0(\omega)$ at $x_0 = 0$, then
\begin{equation}
    \delta \tilde{\varphi}_p(\omega) = \tilde{\varphi}_p(\omega) - \tilde{\varphi}_0(\omega).
    \label{eq:delta_varphi}
\end{equation}
Defining the \textit{phase profile of the metasurface} $\tilde{\varphi}(\bm{r}_p,\omega)$, then at location $\bm{r}_p$, we have $\tilde{\varphi}(\bm{r}_p,\omega) = \tilde{\varphi}_p(\omega)$. By combining \eqref{eq:varphi_d} and \eqref{eq:delta_varphi}, we obtain that for any $\omega$,
\begin{equation}
    \tilde{\varphi}(\bm{r}_p,\omega) = \varphi(\bm{r}_p,\omega) + \tilde{\varphi}_0(\omega).
    \label{eq:tilde_varphi}
\end{equation}
By definition, we can express the \textit{group delay} by
\begin{equation}
    h\frac{d k_p}{d\omega} = \frac{d\tilde{\varphi}_p}{d\omega} = \frac{h}{v_{g,p}},
    \label{eq:group_delay}
\end{equation}
and the \textit{group delay dispersion}
\begin{equation}
    h\frac{d^2 k_p}{d\omega^2} = \frac{d^2\tilde{\varphi}_p}{d\omega^2} = \frac{d}{d\omega}\left[ \frac{h}{v_{g,p}} \right],
    \label{eq:group_delay_disp}
\end{equation}
where $v_{g,p}$ is the group velocity of a pulse propagating in the $p$-th meta-atom.
From the generalized laws of refraction \cite{yu_light_2011}, what is important is the spatial gradient of the target phase profile,
\begin{equation}
    \nabla\varphi(\bm{r},\omega) =  \frac{\omega}{c} \nabla g(\bm{r}).
    \label{eq:gradient}
\end{equation}
Consequently, for any meta-atoms $p$ and $q$, their phase difference is $\tilde{\varphi}_p(\omega) - \tilde{\varphi}_q(\omega) = \frac{\omega}{c} [g(\bm{r}_p) - g(\bm{r}_q)]$.
Taking $q=0$, our reference meta-atom, we obtain the achromaticity condition~\eqref{m-eq:ms_phase_profile} of the main text
\begin{equation}
    \tilde{\varphi}_p(\omega) - \tilde{\varphi}_0(\omega) = \frac{\omega}{c} g(\bm{r}_p).
    \label{eq:condition_disp}
\end{equation}
Fixing a reference meta-atom with phase dispersion $\tilde{\varphi}_0$ imposes the type of dispersion for the achromatic metasurface. We call $\mathcal{D}(\tilde{\varphi}_0)$, \textit{the achromatic subset of meta-atoms associated to the phase dispersion $\tilde{\varphi}_0$.}

Now, taking the successive derivatives of \eqref{eq:condition_disp} with regards to the angular frequency, the difference of group delay between the meta-atoms is given by 
\begin{equation}
    \frac{d \tilde{\varphi}_p}{d\omega}(\omega) - \frac{d \tilde{\varphi}_0}{d\omega}(\omega) = \frac{1}{c} g(\bm{r}_p),
    \label{eq:condition_delay}
\end{equation}
and for any $m\geq 2$, the difference of group delay dispersion is
\begin{equation}
	\frac{d^m \tilde{\varphi}_p}{d\omega^m}(\omega) - \frac{d^m \tilde{\varphi}_0}{d\omega^m}(\omega) = 0 \quad \text{for}\quad m\geq 2.
    \label{eq:condition_delay_disp_n_1}
\end{equation}
Evaluating \eqref{eq:condition_disp}, \eqref{eq:condition_delay} and \eqref{eq:condition_delay_disp_n_1} at the central frequency $\omega_c$, we obtain \eqref{m-eq:achromaticity_cond_a}-(\ref{m-eq:achromaticity_cond_c}) of the main text. Equation~(\ref{eq:condition_delay}) means that a gaussian pulse passing through the center of a metalens should reach the focal point at the same time as a gaussian pulse passing through the edge. To achieve this, it is required to delay the pulse passing through the center of the metasurface relatively to the pulse passing by the edge of the metasurface, i.e increasing the group delay. As the size of the metasurface increases, the necessary delay also increases. It implies a fundamental limit on the maximum group delay photonic structures. Additionally, \eqref{eq:condition_delay_disp_n_1} means that both pulses should undergo the same distortion, i.e keep the same shape.

We would like to comment on the naming the \textit{achromaticity condition} \eqref{m-eq:ms_phase_profile} of the main text. In our definition of the achromatism, the optical function of the metasurface must remain the same for every frequency of the bandwidth. However strictly, \eqref{eq:condition_disp} means that the metasurface has both no chromatic aberrations and no geometrical aberrations. In other words, this equation does not strictly characterize achromaticity. Alternatively, one can derive a more specific condition of achromaticity. Thanks to the linearity in frequency of the target phase profile~(\ref{eq:opd_phase}), it is easy to see that for every meta-atom $p$ of the metasurface and at every frequency $\omega$,
\begin{equation}
    \frac{\Delta\tilde{\varphi}_p(\omega) - \Delta\tilde{\varphi}_p(\omega_\Min)}{\Delta\tilde{\varphi}_p(\omega_\Max) - \Delta\tilde{\varphi}_p(\omega_\Min)} = \frac{\omega - \omega_\Min}{\omega_\Max - \omega_\Min},
    \label{eq:achromaticit_2}
\end{equation}
where we introduce the notation $\Delta\tilde{\varphi}_p = \tilde{\varphi}_p(\omega) - \tilde{\varphi}_0(\omega) = \frac{\omega}{c}g(\bm{r}_p)$. We can see that \eqref{eq:achromaticit_2} is completely independent of the target phase profile. Here, it means that the ratio of phase gradient differences must remain constant over the whole metasurface, which is another way to quantify that the optical function remains identical at every frequency. This is independent of the geometrical aberrations because this condition can remain satisfied even though the metasurface phase profile deviates from the target one. We note that \eqref{eq:achromaticit_2} can serve as an alternative way to define an achromatic subset independently of the notion of target phase profile, as we implicitly did in the main text. However, if we want to design an achromatic metasurface, we want to minimize both aberrations, so considering Eq.\eqref{m-eq:ms_phase_profile} of the main text is necessary.

Defining $n_p$ the effective refractive index of the $p$-th meta-atom of thickness $h$, the instantaneous phase is given by
\begin{equation}
    \tilde{\varphi}_p(\omega) = \frac{\omega}{c} n_p(\omega) h.
    \label{eq:np}
\end{equation}
Here, we emphasize that the \textit{effective} refractive index is not a proper \textit{effective} index in the sense of the \textit{effective medium theory}. This refractive index results from the complex interaction of incident light with a slab composed of identical meta-atoms. The incident light, assumed to be at normal incidence, can couple to several modes of each of these meta-atom and each of these meta-atom can couple to its neighbors. Consequently, the equivalent material can be quite dispersive and the dispersion will depend on (i) the material properties of the meta-atoms, (ii) their geometrical parameters and (iii) the angle of incidence. Applying \eqref{eq:np} to \eqref{eq:condition_delay}, yields for any frequency,
\begin{equation}
    \frac{dn_p}{d\omega} - \frac{dn_0}{d\omega} = 0.
    \label{eq:dndw}
\end{equation}
In other words, to make an achromatic metasurface, all the meta-atoms should have the same index dispersion.

\section{Maximum Size of Achromatic Metasurfaces}\label{sec:size_cond}
We consider a metasurface with an aberrated wavefront,
\begin{equation}
    \tilde{\varphi}_p(\omega)  - \tilde{\varphi}_0(\omega)= \frac{\omega}{c}\gamma(\bm{r}_p,\omega)
    \quad\text{where}\quad
    \gamma(\bm{r},\omega) = g(\bm{r}) + \delta g(\bm{r},\omega),
    \label{eq:phase_metasurface}
\end{equation}
composed of meta-atoms with phase dispersion $\tilde{\varphi}_p$. We suppose that those meta-atoms are  such that over the bandwidth $\Delta\omega$, their transmission is non-zero. Strictly, if the metasurface works in transmission $T$, and has meta-atoms resonant at a certain frequency where $T(\omega_2) = 0$, then the phase delay between the two plane waves is infinity because one is transmitted while the other one is reflected. In practice, when realizing a metasurface operating in transmission, we select meta-atoms with relatively high transmission, avoiding those issues.  Consequently, the time delay between two plane waves of frequencies $\omega_1$ and $\omega_2$ passing by two different meta-atoms $p$ and $q$ is bounded,
\begin{equation}
    \abs{\frac{\tilde{\varphi}_q(\omega_2)}{\omega_2} - \frac{\tilde{\varphi}_p(\omega_1)}{\omega_1}} \leq \Delta\tau,
    \label{eq:bound_delta_t}
\end{equation}
with $\Delta\tau$ a certain bound. Applying \eqref{eq:phase_metasurface} to \eqref{eq:bound_delta_t}, it follows that,
\begin{equation}
    \abs{\frac{1}{c}\gamma(\bm{r}_q,\omega_2) - \frac{1}{c}\gamma(\bm{r}_p,\omega_1) + \frac{\tilde{\varphi}_0(\omega_2)}{\omega_2} - \frac{\tilde{\varphi}_0(\omega_1)}{\omega_1}} \leq \Delta\tau.
    \label{eq:bound_delta_t_2}
\end{equation}

We can distinguish three cases, (i) the case with $p = q$ and $\omega_1 \neq \omega_2$, (ii) when $p \neq q$ and $\omega_1 = \omega_2$ and (iii) $p \neq q$ and $\omega_1 \neq \omega_2$.

\textbf{If $p = q$ and $\omega_1 \neq \omega_2$}.
\begin{equation}
    \abs{\frac{\tilde{\varphi}_0(\omega_2)}{\omega_2} - \frac{\tilde{\varphi}_0(\omega_1)}{\omega_1}} \leq \Delta\tau.
    \label{eq:bound_delta_t_omega}
\end{equation}
This equation quantifies the total phase delay, over the bandwidth, a physical system can impose on pulses as in slow-light delay lines \cite{miller_fundamental_2007}.

\textbf{If $p\neq q$, and $\omega_1 = \omega_2 = \omega$}.
\begin{equation}
    \abs{\gamma(\bm{r}_q,\omega) - \gamma(\bm{r}_p,\omega)} \leq c \Delta\tau.
    \label{eq:bound_delta_t_pq}
\end{equation}
This is the monochromatic size limit at frequency $\omega$ without phase wrapping.

\textbf{If $p\neq q$, and $\omega_1 \neq \omega_2$}. This case is particular to the case of achromatic metasurface as opposed to the two previous conditions. Because \eqref{eq:bound_delta_t_2} is true for any couple of frequencies $(\omega_1, \omega_2)$, this is in particular true for the couple of frequencies such that $\abs{\frac{\tilde{\varphi}_0(\omega_2)}{\omega_2} - \frac{\tilde{\varphi}_0(\omega_1)}{\omega_1}}$ is maximum. In most practical case, this is the case when $\omega_2 = \omega_\Max$ and $\omega_1 = \omega_\Min$. We define
\begin{equation}
    \Delta\tau_0 =\frac{\tilde{\varphi}_0(\omega_\Max)}{\omega_\Max} - \frac{\tilde{\varphi}_0(\omega_\Min)}{\omega_\Min}.
    \label{eq:delta_t_0}
\end{equation}
Similarly, \eqref{eq:bound_delta_t_2} is valid for any meta-atoms, so it is true for the couple of meta-atoms such that $\abs{\gamma(\bm{r}_q,\omega_\Max) - \gamma(\bm{r}_p,\omega_\Min)}$ is maximal. We note,
\begin{equation}
    \Delta g + \Delta\delta g = \max_{p,q} \abs{\gamma(\bm{r}_q,\omega_\Max) - \gamma(\bm{r}_p,\omega_\Min)} ,
\end{equation}
leading to
\begin{equation}
    \frac{\Delta g}{c} + \frac{\Delta\delta g}{c} + \Delta\tau_0 \leq \Delta \tau.
    \label{eq:size_cond_dg}
\end{equation}
Equation~\eqref{eq:size_cond_dg} is slightly different from \eqref{m-eq:size_cond} of the main text. Here, we have an additional term $\Delta\delta g$ accounting for the wavefront aberrations with regards to the target optical function. Thus the term $\Delta\delta g$ can be positive or negative depending on how the aberrations affects the profile of the metasurface. In the case of the manuscript, meta-atoms belong to an achromatic subset, so this term disappears, i.e. $\Delta\delta g = 0$. We emphasize that the reciprocal is not necessarily true, and $\Delta\delta g = 0$ does not mean the Strehl ratio of the metasurface will be unitary.

Regarding the size of the metasurface in presence of aberrations, it cannot be directly determined by inversion of $g$ as it depends on $\Delta\delta g$ which is an unknown quantity. Positive values of $\Delta\delta g$ decrease the size of the achievable metasurface compared to a perfectly achromatic metasurface spanning the full range of $\Delta\tau - \Delta\tau_0$, while negative values increase it. This additional third term in \eqref{eq:size_cond_dg} is the reason why we obtain the size-bandwidth trade-off condition. Indeed, this term is function of the bandwidth $\Delta \omega$, meaning that for different values of the bandwidth, we would have different bounds for $\Delta g$. However, from \eqref{eq:size_cond_dg} it remains challenging to derive a trade-off condition for size-bandwidth-optical performances, without taking particular examples. 

\section{Bandwidth independence on achromatic subset}\label{sec:bandwidth_achromatic}
We consider meta-atoms belonging to an achromatic subset $\mathcal{D}(\tilde{\varphi}_0)$. According to \eqref{m-eq:size_cond} of the main text, they satisfy
\begin{equation}
    \frac{\Delta g}{c} \leq \Delta\tau - \Delta\tau_0
\end{equation}
where $\Delta\tau = \max_{(p,q),(\omega_1,\omega_2)}\abs{\frac{\tilde{\varphi}_q(\omega_2)}{\omega_2} - \frac{\tilde{\varphi}_p(\omega_1)}{\omega_1}}$ and $\Delta \tau_0 = \max_{\omega_1,\omega_2} \abs{\frac{\tilde{\varphi}_0(\omega_2)}{\omega_2} - \frac{\tilde{\varphi}_0(\omega_1)}{\omega_1}}$.

In general, the phase delay is monotonically increasing with the frequency thus we assume the minimum and maximum value are obtained at the border of the bandwidth, leading to
\begin{equation}
    \Delta\tau_0 = \frac{\tilde{\varphi}_0(\omega_\Max)}{\omega_\Max} - \frac{\tilde{\varphi}_0(\omega_\Min)}{\omega_\Min},
    \label{eq:delta_tau_0}
\end{equation}
and
\begin{equation}
    \Delta\tau = \frac{\tilde{\varphi}_p(\omega_\Max)}{\omega_\Max} - \frac{\tilde{\varphi}_q(\omega_\Min)}{\omega_\Min},
    \label{eq:delta_tau}
\end{equation}
with $p$ and $q$ the indices of the meta-atoms corresponding to the maximum and minimum value of phase delay.
Without loss of generality, we can safely take $q = 0$, \textit{i.e.} the reference meta-atom. By taking the difference of \eqref{eq:delta_tau} by \eqref{eq:delta_tau_0}, we obtain that
\begin{equation}
    \Delta\tau - \Delta\tau_0 = \frac{\tilde{\varphi}_p(\omega_\Max) - \tilde{\varphi}_0(\omega_\Max)}{\omega_\Max},
\end{equation}
where
\begin{equation}
    \tilde{\varphi}_p(\omega_\Max) - \tilde{\varphi}_0(\omega_\Max) = \Delta\beta_{p,0} + \frac{\Delta\omega}{2}\Delta\beta_{p,1} + \frac{\Delta\omega^2}{8}\Delta\beta_{p,2},
\end{equation}
Since the meta-atom $p$ belongs to the achromatic subset $\mathcal{D}(\tilde{\varphi}_0)$, we can apply \eqref{m-eq:achromaticity_cond_a}-(\ref{m-eq:achromaticity_cond_c}) of the main text, leading to
\begin{equation}
    \tilde{\varphi}_p(\omega_\Max) - \tilde{\varphi}_0(\omega_\Max) = \left(\omega_c + \frac{\Delta\omega}{2}\right) \Delta\beta_{p,1} = \omega_\Max \Delta\beta_{p,1},
\end{equation}
It naturally follows that
\begin{equation}
    \Delta\tau - \Delta\tau_0 = \Delta\beta_{p,1},
\end{equation}
which is independent of the bandwidth.

\section{Strehl Ratio Expansion in Chromatic Aberrations}\label{sec:strehl_ratio}
Considering a bandwidth $\Delta\omega$ centered in $\omega_c$, we approximate the phase dispersion $\tilde{\varphi}_p$ by its Taylor expansion in the neighboring $\delta\omega$ of the central frequency $\omega_c$,
\begin{equation}
        \tilde{\varphi}_p(\omega) = P_p^{(M)}(\omega) + o\left((\omega-\omega_c)^m\right)
        \quad\text{with}\quad
        P_p^{(M)}(\omega) = \sum_{m=0}^M \frac{(\omega-\omega_c)^m}{m!}  \left.\frac{d^m\tilde{\varphi}_p}{d\omega^m}\right\rvert_{\omega_c}
        \label{eq:taylor_expansion}
\end{equation}
If we realize a metasurface with those meta-atoms, then, the real phase profile of the metasurface is given by
\begin{equation}
    \tilde{\varphi}(\bm{r}_p, \omega) =  \sum_{m=0}^M \frac{(\omega-\omega_c)^m}{m!}  \Delta\beta_{p,m},
    \quad\text{with}\quad
    \Delta\beta_{p,m} = \beta_{p,m} - \beta_{0,m}
    \quad\text{and}\quad
    \beta_{p,m} = \left.\frac{d^m\tilde{\varphi}_p}{d\omega^m}\right\rvert_{\omega_c}.
    \label{eq:err_fct_condition_approx}
\end{equation}
Evidently, for any meta-atom $p$ and at the central frequency $\omega_c$, $\tilde{\varphi}(\bm{r}_p, \omega_c) =  \varphi(\bm{r}_p,\omega_c)$, i.e. the metasurface is designed to operate at the central frequency in a monochromatic way. We define the phase aberration
\begin{equation}
    \delta\varphi_p(\omega) = \varphi(\bm{r}_p,\omega) - \tilde{\varphi}(\bm{r}_p,\omega).
\end{equation}
Using \eqref{eq:opd_phase},
\begin{equation}
    \delta\varphi_p(\omega) = \frac{\omega}{c} g(\bm{r}_p) - \sum_{m=0}^M \frac{(\omega-\omega_c)^m}{m!}  \Delta\beta_{p,m}.
\end{equation}
Adding and substracting the term $\frac{\omega_c}{c}g(\bm{r}_p)$,
\begin{equation}
    \delta\varphi_p(\omega) = \frac{\omega_c}{c} g(\bm{r}_p) - \Delta\beta_{p,0} + \frac{(\omega - \omega_c)}{c} g(\bm{r}_p) - \sum_{m=1}^M \frac{(\omega-\omega_c)^m}{m!}  \Delta\beta_{p,m}.
\end{equation}
Leading to,
\begin{equation}
    \delta\varphi_p(\omega) =  \sum_{m=0}^M \frac{(\omega-\omega_c)^{m}}{m!}  \delta_{p,m},
\end{equation}
with
\begin{equation}
    \delta_{p,0} = \frac{\omega_c}{c}g(\bm{r}_p) - \Delta\beta_{p,0},
    \quad
    \delta_{p,1} = \frac{g(\bm{r}_p)}{c} - \Delta\beta_{p,1} 
    \quad\text{and}\quad
    \delta_{p,m} = - \Delta\beta_{p,m},
\end{equation}
corresponding to the zeroth, first and higher order chromatic aberrations of the wavefront.

Since the metasurface is composed of a discrete array of meta-atoms, we approximate the Strehl ratio, assuming a 2D array of $N$ meta-atoms on a square lattice with period $\Lambda$,
\begin{equation}
    S(\omega) \approx \frac{\Lambda^4}{\pi^2 a^4} \abs{\sum_p e^{i \delta\varphi_p(\omega) }}^2.
    \label{eq:strehl_summation}
\end{equation}
Considering the aberrations are small enough, we have
\begin{align}
    e^{i \delta\varphi_p(\omega) } \approx 1
    + i\delta\varphi_p - \frac{1}{2}\delta\varphi_p^2 - \frac{i}{6}\delta\varphi_p^3 + \frac{1}{24}\delta\varphi_p^4.
    \label{eq:aberrations_order}
\end{align}
Posing $x = \omega-\omega_c$, we consider only up to the second order chromatic aberration on the wavefront,
\begin{equation}
    \delta\varphi_p(\omega) \approx \delta_{p,0} + x \delta_{p,1} + \frac{x^2}{2}\delta_{p,2}.
    \label{eq:aberrations_hyp}
\end{equation}
We insert \eqref{eq:aberrations_hyp} into \eqref{eq:aberrations_order}. We keep only up to the power 2 of $\delta_p$. For the real part, we obtain 
\begin{equation}
    \real\left[e^{i\delta\varphi_p}\right] \approx 1 - \frac{1}{2} \delta_{p,0}^2 - x \delta_{p,0}\delta_{p,1} - \frac{1}{2} x^2 (\delta_{p,1}^2 + \delta_{p,0}\delta_{p,2}) - \frac{1}{2} x^3 \delta_{p,1} \delta_{p,2} - \frac{x^4}{8} \delta_{p,2}^2,
\end{equation}
and for the imaginary part
\begin{equation}
    \imag\left[e^{i\delta\varphi_p}\right] \approx \delta_{p,0} + x \delta_{p,1} + \frac{x^2}{2}  \delta_{p,2}.
\end{equation}
Thus by taking the real part of the summation over all meta-elements $p$ we obtain
\begin{equation}
    \frac{1}{N} \sum_p \real[e^{i\delta\varphi_p}] \approx 1 - \frac{1}{2}\braket{\delta_0^2} - x \braket{\delta_0\delta_1} - \frac{1}{2} x^2 \left[ \braket{\delta_1^2} + \braket{\delta_0\delta_2} \right] - \frac{1}{2} x^3 \braket{\delta_1\delta_2} - \frac{1}{8} x^4 \braket{\delta_2^2},
\end{equation}
and
\begin{equation}
    \frac{1}{N} \sum_p \imag[e^{i\delta\varphi_p}] \approx  \braket{\delta_0} + x \braket{\delta_1} + \frac{1}{2} x^2 \braket{\delta_2},
\end{equation}
where we have defined the notation
\begin{equation}
    \braket{\delta_m\delta_n} = \frac{1}{N} \sum_p \delta_{p,m}\delta_{p,n}.
\end{equation}
The Strehl ratio is thus obtained by taking the square of the absolute value of the summation,
\begin{equation}
    S(\omega) = \abs{\sum_p \real[e^{i\delta\varphi_p}]}^2 + \abs{\sum_p \imag[e^{i\delta\varphi_p}]}^2.
\end{equation}
Consequently, we obtain an approximation of the Strehl ratio in function of the aberrated wavefront,
\begin{equation}
    S(\omega) \approx \frac{N^2\Lambda^4}{\pi^2 a^4}\left[ 1 - V_{0,0} - 2(\omega-\omega_c) V_{0,1} - (\omega-\omega_c)^2 \left( V_{1,1} + V_{0,2} \right) - (\omega-\omega_c)^3 V_{1,2} - \frac{(\omega-\omega_c)^4}{4} V_{2,2} \right],
    \label{eq:strehl_expansion}
\end{equation}
with $V_{m,n}$ the covariance between the chromatic aberration of the wavefront of orders $m$ and $n$,
\begin{equation}
    V_{m,n} = \braket{\delta_m\delta_n} - \braket{\delta_m}\braket{\delta_n}.
    \label{eq:V_pq}
\end{equation}
We can make several remarks
\begin{itemize}
    \item The quantity $N \Lambda^2$ corresponds to the surface of the sampled metasurface and $\pi a^2$ the surface of the aperture, thus its ratio represents the packing density of meta-atoms inside the metasurface. The number $N$ depends on the arrangement of the meta-atoms.
    \item At $\omega = \omega_c$, the Strehl ratio is exactly the standard expression of the Strehl ratio as a function of the variance of the phase error.
    \item The variance of the first and second order aberrations appears respectively in the terms $(\omega - \omega_c)^2$ and $(\omega - \omega_c)^4$. Consequently, for small bandwidth, those aberrations do not play an important role.
    \item The terms in $(\omega - \omega_c)$ and $(\omega - \omega_c)^3$ are associated with the covariance between the aberrations of different orders.
\end{itemize}

 The covariance between two quantities quantifies some relationship between them. It quantifies the average of the product of the difference between two variables with their respective mean,
 \begin{equation}
    V_{m,n} = \braket{\left(\delta_{p,m} - \braket{\delta_n}\right) \left(\delta_{p,m} - \braket{\delta_n}\right)}.
    \label{eq:V_pq_2}
\end{equation}
 If the covariance is positive, then when the value of one quantity is large, the other is large too. On the contrary, if the covariance is negative, then when the value of one quantity is large, the other tends to be small. Considering a meta-atom $p$ such that $\delta_{p,0} > \braket{\delta_0}$, then a monochromatic plane wave at frequency $\omega_c$ would arrive at a later time at the focal point of the metalens compared to the same monochromatic plane wave transmitted through the rest of the meta-atoms. Now, if $V_{0,1} > 0$, it would mean that a gaussian pulse transmitted by the meta-atom $p$ would also arrive at a later time. Reversely, if $V_{0,1} < 0$, then the gaussian pulse would arrive earlier than for the rest of the meta-atoms.

\section{Meta-atoms Selection Method Comparison}\label{sec:selection_strategies}
\begin{figure}
    \centering
    \includegraphics[width=\textwidth]{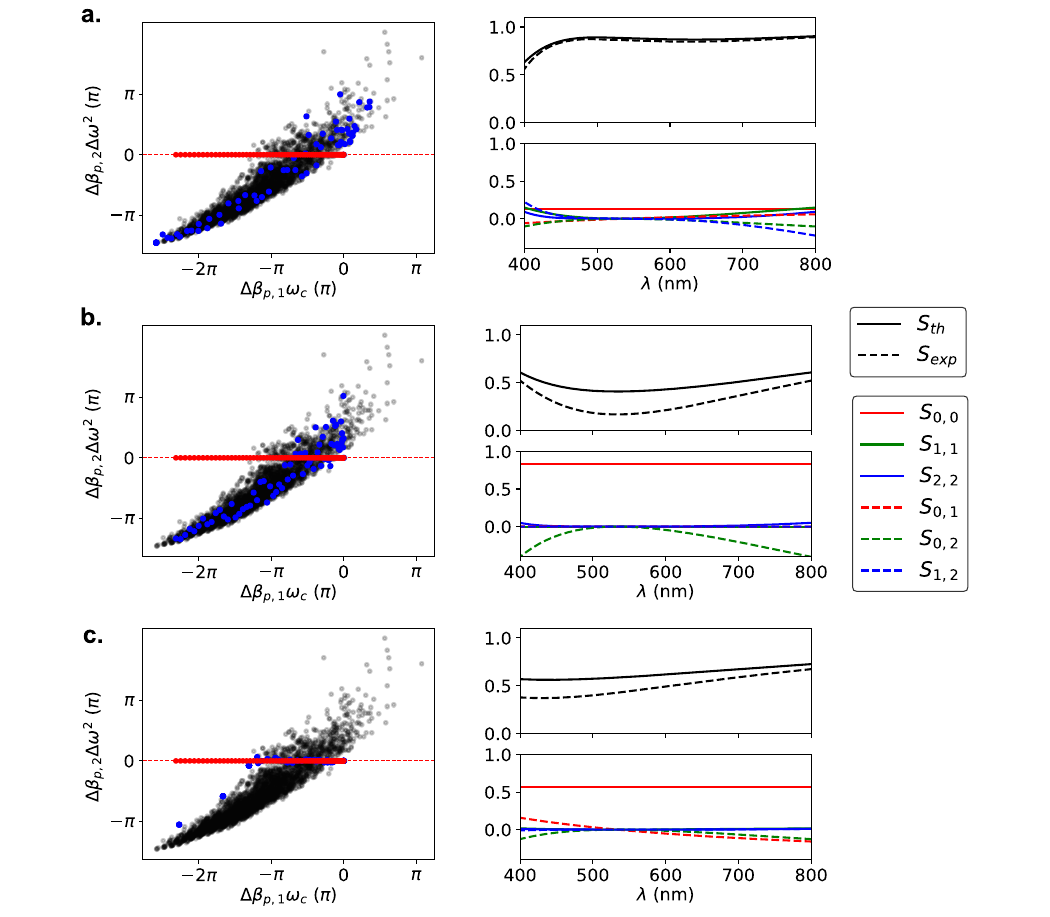}
    \caption{\textbf{a.} and \textbf{b.} Scenario 1: Monochromatic Design.\textbf{c.} and  \textbf{d.} Scenario 2: Group Delay Optimization.  In each case, \textbf{a.} and \textbf{c.} represent the dataset (black points), the selected meta-atoms (blue points) and the target metalens profile (red points). \textbf{b.} and \textbf{d.} represent the Strehl ratio. (Top panels) Comparison between the theoretical Strehl ratio, $S_{th}$ obtained with \eqref{eq:strehl_summation} and the Strehl ratio expansion, $S_{exp}$, obtained with \eqref{eq:strehl_expansion}. (Bottom panels) Components of the Strehl ratio expansions.}
    \label{fig:sm_strehl_expansion}
\end{figure}

In the main manuscript, we choose the meta-atoms in order to minimize the difference between the target phase profile and the phase dispersion of the meta-atoms in terms of the infinite norm, $\norm{\varphi - \tilde{\varphi}_p}_{\infty}$. Here, we consider three alternate cases, (i) minimizing the error on the phase profile at the central frequency; (ii) minimizing the first order aberrations; (iii) minimizing simultaneously the three aberrations orders by means of Euclidean distances in phase space. The metasurface is the same as in the main manuscript with numerical aperture $\text{NA} = 0,05$ and diameter \SI{50}{\micro\metre}. In Fig.~\ref{fig:sm_strehl_expansion}, we show the computation of the Strehl ratio~\eqref{eq:strehl_expansion} and compare it with the exact value obtained by \eqref{eq:strehl_summation}, for all scenarii, where
\begin{equation}
    S(\omega) \approx 1 - S_{0,0} - S_{0,1} - S_{1,1} - S_{0,2} - S_{1,2} - S_{2,2}.
    \label{eq:strehl_expansion_2}
\end{equation}

\textbf{Scenario 1. Optimization of the phase at the central frequency.} The variance $V_{0,0}$ is null, and the covariances $V_{0,1}$ and $V_{0,2}$ are likely negligible thus,
\begin{equation}
    S(\omega) \approx  1 - S_{1,1} - S_{1,2} - S_{2,2},
\end{equation}
In Fig.~\ref{fig:sm_strehl_expansion} (a), the metasurface is not optimal at the central frequency because the required $\Delta\beta_{p,0}$ range exceeds the available $\Delta\beta_{p,0}$ range as shown in the main text, Fig. 4 (b). Thus the 0-th order aberration is quite large with $S_{0,0} \approx 0.18$ and so we have non-negligible terms $S_{0,1}$, and $S_{0,2}$. But overall, the Strehl ratio is acceptably good except at small wavelengths where it dips at $0.5$. This can be explained by looking at $S_{1,1}$ and $S_{2,2}$ corresponding to the variance of the first and second order chromatic aberrations of the wavefront, which are both positive and thus, contribute to reduce the overall performance, especially at small wavelengths. On the other hand the overall good behavior can be explained by looking at the covariance terms $S_{0,2}$ and $S_{1,2}$ which are negative. In other words, despite not having the exact behavior required, in terms of group delay and group delay dispersion, the errors are compensated by the fact that a pulse would arrive slightly later or with a shape slightly increased.

\textbf{Scenario 2. Optimization of the group delay only.} The variance $V_{1,1}$ is null, and the covariances $V_{0,1}$ and $V_{1,2}$ are likely negligible, cf. Fig.~\ref{fig:sm_strehl_expansion} (b),
\begin{equation}
    S(\omega) \approx 1 - S_{0,0} - S_{0,2} - S_{2,2}.
\end{equation}
Because we did not optimize the 0-th order aberrations, then the term $S_{0,0}$ is larger than in previous case. The group delay dispersion being also not optimized, the term $S_{2,2}$ reduce the overall value of the Strehl ratio. However, the terms $S_{0,2}$ is largely negative. For large value of the phase profile, the group delay dispersion is such that the deformation of the pulse compensate for the aberrations in 0-th order.

\textbf{Scenario 3. Euclidean distance in phase space.} In this scenario, we choose the meta-atoms in order to minimize simultaneously $V_{0,0}$, $V_{1,1}$ and $V_{2,2}$ by minimizing the Euclidean distance in phase space cf. Fig.~\ref{fig:sm_strehl_expansion} (c). If those three terms are small then the Strehl ratio should be unitary and the metasurface would be achromatic as per the achromaticity condition. However, in this case the phase profile at the central frequency cannot match perfectly the target phase profile, thus $S_{0,0}$ is non-zero, leading to
\begin{equation}
    S(\omega) \approx 1 - S_{0,0} - S_{0,1} - S_{0,2}.
\end{equation}
It is remarkable that despite minimizing all parameters, we obtain worse optical performances than simply designing a monochromatic metasurface. The reason lies essentially in having a strong geometrical aberrations, $S_{0,0} > 0.5$, meaning that the metasurface barely achieve the optical function it has been designed to do. By trying to minimize the Euclidean distance in phase space, we attribute the same importance to the different orders of chromatic aberrations of the wavefront, while Eq.~\eqref{eq:strehl_expansion} clearly shows that the $V_{m,n}$ terms have different impacts on the Strehl ratio value.

\section{Figure of Merit: Mean Strehl Ratio}\label{sec:mean_strehl}
We define by $\braket{S}$ the mean of the Strehl ratio over the bandwidth,
\begin{equation}
    \bar{S} = \frac{1}{\Delta\omega} \int_{\omega_\Min}^{\omega_\Max} S(\omega) d\omega.
    \label{eq:strehl_mean}
\end{equation}
leading to
\begin{equation}
    \bar{S} = 1 - V_{0,0} - \frac{V_{1,1} + V_{0,2}}{12} \Delta\omega^2 - \frac{V_{2,2}}{320} \Delta\omega^4.
    \label{eq:strehl_mean_2}
\end{equation}
The covariance terms $V_{0,1}$ and $V_{1,2}$ do not contribute in this approximation to the average Strehl ratio over the bandwidth.

\subsection{Non-unitary Strehl ratio achromatic metasurface}
According to \eqref{eq:strehl_mean_2}, if we consider meta-atoms that matches the group delay and group delay dispersion but not the instantaneous phase at the central frequency, then 
\begin{equation}
    \bar{S} = 1 - V_{0,0}.
\end{equation}
The metasurface is achromatic but with an aberrated phase profile. Therefore to induce chromatic aberrations we need to consider that either the condition on the group delay or group delay dispersion is not satisfied.

\subsection{Quasi-achromatic Metasurface}
We consider a metasurface designed to be achromatic except the last meta-atoms on the outer boundary. Since the last meta-atom $p_0$ is not on the achromatic subset, we suppose only the condition on the group delay is not satisfied,
\begin{equation}
    \left\{
    \begin{aligned}
        &\Delta\beta_{p_0,0} = \frac{\omega_c}{c} g(r_{p_0})\\
        &\Delta\beta_{p_0,1} \neq \frac{1}{c} g(r_{p_0})\\
        &\Delta\beta_{p_0,2} = 0
    \end{aligned}
    \right. .
\end{equation}
Thus the Strehl ratio can be approximated by
\begin{equation}
    \bar{S} = 1 - \Delta\omega^2 \frac{V_{1,1}}{12}.
\end{equation}
This equation is dependent on the variance of the $1$-st order chromatic aberrations $V_{1,1}$ which are unknown and depend on the choice of the meta-atoms. We assume we make an error $\varepsilon(R)$, proportional to the expected value, at the position $R$,
\begin{equation}
    \delta_{p,1} = \left\{
    \begin{aligned}
        &\varepsilon(R) \quad\text{if}\quad p = p_0\\
        &0 \quad\text{otherwise}
    \end{aligned}
    \right. ,
\end{equation}
Considering we have $N$ meta-atoms on the metasurface and $N_0$ on the outer perimeter of the metasurface, from \eqref{eq:V_pq},
\begin{equation}
    V_{1,1} = \varepsilon^2 \left(\frac{N_0}{N} - \frac{N_0^2}{N^2}\right) \approx \varepsilon^2 \frac{N_0}{N}.
\end{equation}
So we have
\begin{equation}
    \bar{S} = 1 - \Delta\omega^2 \frac{N_0}{N} \frac{\varepsilon^2}{12}.
\end{equation}
To get an estimation of the Strehl ratio, we need to make some assumptions on the evolution of $\varepsilon$ in function of the parameters. In other words, we need to parametrize the variance $V_{1,1}$. Each parametrization leads to a relationship between $\bar{S}, \Delta\omega$ and $D$ but it has to be interpreted with regards the initial assumptions leading to this parametrization.

In the following, the meta-atoms are placed in square grid, so we have approximately with period $p$,
\begin{equation}
    N \approx \frac{\pi R^2}{p^2}, \quad\text{and}\quad N_0 \approx \frac{2\pi R}{p},
\end{equation}

\textbf{Parametrization 1. Constant $\varepsilon$.}
If we fix the error $\varepsilon$ to be a constant, then it means that no matter the size of the metasurface, the error on the group delay of the outer rings of meta-atoms remains the same, leading to
\begin{equation}
    \frac{\Delta\omega^2}{D (1 - \bar{S})} = \frac{3}{\varepsilon^2 p}
\end{equation}
This equation seems counter-intuitive since for larger metasurfaces, we can realize larger bandwidths at constant Strehl ratio and constant aberrations. It comes from the fact that the quantity $\varepsilon$ quantifies an absolute error on the group delay, meaning that smaller metasurfaces are more aberrated compared to their larger counter parts. This result is also independent of the type of metasurface we are realizing.

\textbf{Parametrization 2. Relative error $\varepsilon(R) = \alpha \frac{g(R)}{c}$.}
The error $\varepsilon$ is dependent on the radius of the metasurface, meaning that outer ring of meta-atoms make an error $\alpha$ on their expected value $g(R)/c$,
\begin{equation}
    \varepsilon(R) = \alpha \frac{g(R)}{c}.
\end{equation}
Thus it follows that $V_{1,1} = 2\alpha^2 \frac{p}{R}\frac{g(R)^2}{c^2}$ and so
\begin{equation}
    \bar{S} = 1 - \Delta\omega^2 \frac{\alpha^2}{6 c^2} \frac{p}{R} g(R)^2.
\end{equation}
We obtain a relationship that gives the average Strehl ratio of the metasurface, for a given bandwidth $\Delta\omega$, diameter of the achromatic metasurface $D = 2R$ and first order chromatic aberrations $\varepsilon$. Here the diameter corresponds to the diameter of the achromatic metasurface plus one meta-atom, i.e $p \ll D$. If this last meta-atom is part of the achromatic dataset, then $\alpha = 0$ and $\bar{S} = 1$, however, as soon as the meta-atoms induces first order chromatic aberrations, the size of the metasurface will impose define bandwidth $\Delta \omega$.

Considering a metalens, $g(R) = f - \sqrt{f^2 + R^2}$, we can express the previous equation in terms of numerical aperture NA,
\begin{equation}
    g(R) = \frac{R}{\text{NA}} \left( \sqrt{1 - \text{NA}^2} - 1\right),
\end{equation}
leading to
\begin{equation}
    \frac{\Delta\omega^2 D}{1 - \bar{S}} = \frac{12 c^2}{p \alpha^2} \frac{\text{NA}^2}{\left( 1 - \sqrt{1 - \text{NA}^2}\right)^2},
\end{equation}
and by taking the square root,
\begin{equation}
    \frac{\Delta\omega \sqrt{D}}{\sqrt{1 - \bar{S}}} = \frac{2\sqrt{3}}{\alpha} \frac{c}{\sqrt{p}} \frac{\text{NA}}{1 - \sqrt{1 - \text{NA}^2}}.
\end{equation}

\textbf{Parametrization 3. Constant standard deviation $V_{1,1} = \alpha^2 \frac{g(R)^2}{c^2}$.}
In the previous example, the variance of the aberrations $V_{1,1}$ depends both on the expected value group delay $g(R)/c$ and on the size $1/R$. We can make another parametrization such that $V_{1,1}$ depends only on $g(R)/c$,
\begin{equation}
    V_{1,1} = \alpha^2 \frac{g(R)^2}{c^2},
\end{equation}
meaning that the group delay of meta-atoms scales following to the relationship
\begin{equation}
    \varepsilon(R) = \alpha \sqrt{\frac{N}{N_0}} \frac{g(R)}{c}.
\end{equation}
In other words, we parametrized the standard deviation $\sqrt{V_{1,1}}$ to remain consistent between metasurfaces of different size. Therefore,
\begin{equation}
    \frac{\Delta\omega D}{\sqrt{1 - \bar{S}}} = 4\sqrt{3} \frac{c}{\alpha} \frac{\text{NA}}{1 - \sqrt{1 - \text{NA}^2}}.
    \label{eq:param_3_bandwidth}
\end{equation}
For a fixed Strehl ratio and numerical aperture NA, the size of the metasurface will evolve as an inverse law of the bandwidth. We can compare the equivalent equation obtained from ref.~\citep{presutti_focusing_2020}, where the bound of the size of the achromatic metasurface is given by
\begin{equation}
    \Delta\omega D \leq \kappa \frac{ \text{NA}}{ 1 - \sqrt{1 - \text{NA}^2}}.
    \label{eq:presutti}
\end{equation}
We can observe that both equations are similar except the presence of the term $1-\bar{S}$ in our expression which appears due to the presence of chromatic aberrations.

\textbf{Parametrization 4. Linear standard deviation $\sqrt{V_{1,1}} = \alpha \frac{N}{N_0} \frac{g(R)}{c}$.}
In the previous example, the variance of the aberrations $V_{1,1}$ depends both on the expected value group delay $g(R)/c$ and on the size $1/R$. We can make another parametrization such that $V_{1,1}$ depends only on $g(R)/c$,
\begin{equation}
    V_{1,1} = \alpha^2 \frac{N^2}{N_0^2} \frac{g(R)^2}{c^2},
\end{equation}
In other words, we parametrized the standard deviation $\sqrt{V_{1,1}}$ to remain consistent between metasurfaces of different size. Therefore,
\begin{equation}
    \frac{\Delta\omega D^2}{\sqrt{1 - \bar{S}}} = 16 \sqrt{3} \frac{pc}{\alpha} \frac{\text{NA}}{1 - \sqrt{1 - \text{NA}^2}}.
    \label{eq:param_4_bandwidth}
\end{equation}

\begin{figure}
    \centering
    \includegraphics[width=\textwidth]{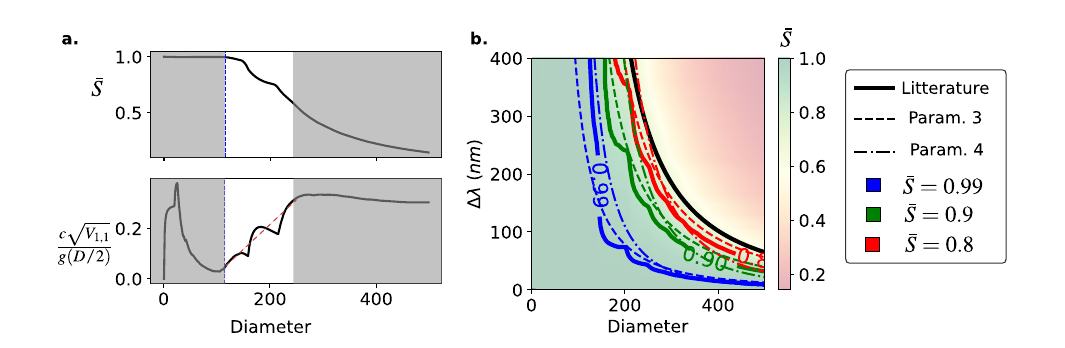}
    \caption{\textbf{a.} Evolution of the ratio $\sqrt{V_{1,1}}$ by $g(D/2)/c$ as a function of the diameter $D$ in the pseudo achromatic regime. \textbf{b.} Strehl ratio as a function of the diameter and bandwidth. The thick color lines represent the iso-Strehl contours. The dashed lines and dashed-dotted lines represent respectively the iso-Strehl ratio computed with parametrization 3, i.e. \eqref{eq:param_3_bandwidth} with $\alpha = 0.2$, and parametrization 4, i.e. Eq.\eqref{eq:param_4_bandwidth}. The black thick line represents the size-bandwidth limit associated with \eqref{eq:presutti}.}
    \label{fig:bandwidth_limit}
\end{figure}

In Fig.~\ref{fig:bandwidth_limit}, we test the parametrization 3 and 4, for the metalens presented in the main text. In Fig.~\ref{fig:bandwidth_limit} (a), we plot the ratio of the standard deviation $\sqrt{V_{1,1}}$ with the $g(D/2)/c$ as a function of the diameter $D$. In the pseudo-achromatic regime, this ratio is approximately increasing linearly with the diameter as in parametrization 4. In Fig.~\ref{fig:bandwidth_limit}(b) we plot the corresponding iso-Strehl ratio contours in the case of parametrization 3 and 4 and show that both formula shows qualitative agreement with the theoretical values. Besides we compare with \eqref{eq:presutti}, which is a larger bound. We emphasize that the comparison is only qualitative, because in our parametrization we neglected the geometrical aberrations and we show in the main text that those aberrations are non-negligible.

\section{Effect of the choice of the norm}\label{sec:norm_choice}

\begin{figure*}
    \centering
    \includegraphics[width=\textwidth]{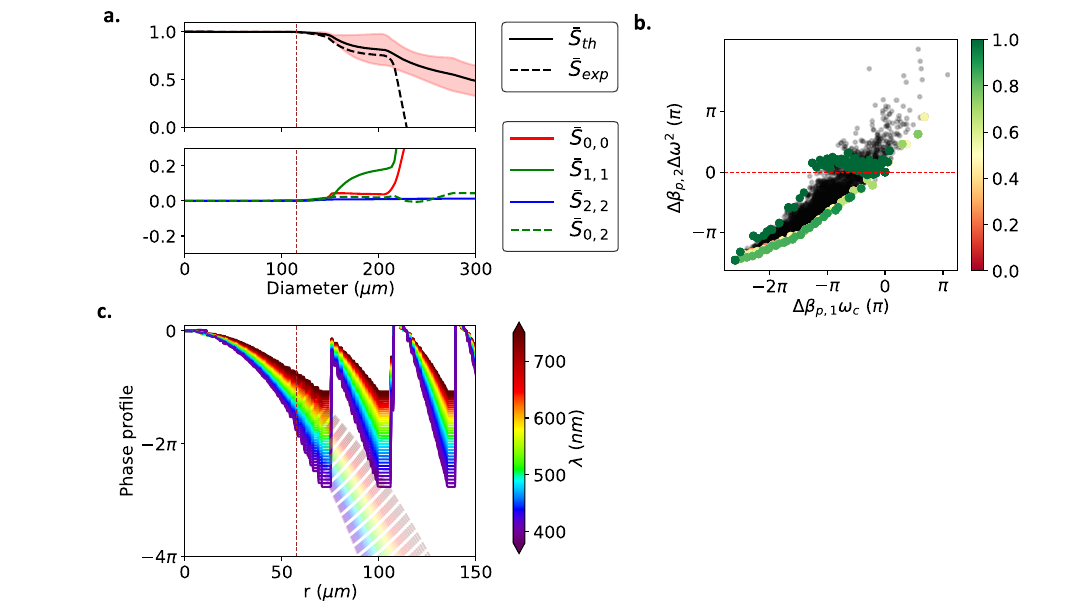}
    \caption[Minimizing $\norm{e^{i\varphi} - e^{i\tilde{\varphi}_p}}_2$]{\textbf{Minimizing $\norm{e^{i\varphi} - e^{i\tilde{\varphi}_p}}_2$} (a) Evolution of the Strehl ratio as a function of the diameter of a metalens with focal length $f = \SI{5}{\milli\metre}$. (Top panel) The black plain curve represents the exact Strehl ratio and the dashed line its expansion approximation. The red area represents the standard deviation of the Strehl ratio $S(\omega)$. (Bottom panel) Strehl ratio expansion coefficient with $\bar{S}_{0,0} = V_{0,0}$, $\bar{S}_{1,1} = \frac{\Delta\omega^2}{12}V_{1,1}$, $\bar{S}_{0,2} = \frac{\Delta\omega^2}{12}V_{0,2}$, and $\bar{S}_{2,2} = \frac{\Delta\omega^4}{320}V_{2,2}$. (b) Representation in phase space of the selected meta-atoms of the dataset with color code representing the average Strehl ratio $\bar{S}$. (c) Evolution of the phase profile of the metasurface as a function of the position $\bm{r}_p$ at each frequency. The plain color lines represent the realized phase profile, and the dashed lines represent the target phase profile. The color map represents each wavelength.}
	\label{fig:norm_choice}
\end{figure*}

In the main manuscript, we minimize the distance $\norm{e^{i\varphi} - e^{i\tilde{\varphi}_p}}_\infty$ and in Fig.~\ref{m-fig:figure_4}(c) we can see that from a certain aperture the phase profile appears to be chaotic. This is because we are minimizing the maximum distance between values that are located on the unit circle of the complex plane. Eventually, for large aperture metalenses, the maximum distance reaches $2$, the diameter of the unit circle and any meta-atoms of the dataset.

We can overcome this issue by minimizing the least square distance instead, $\norm{e^{i\varphi} - e^{i\tilde{\varphi}_p}}_2$, see Fig.~\ref{fig:norm_choice}. The phase profile is no more chaotic and defines clear Fresnel zones.

\bibliography{ref}